\documentclass[aps,prd,twocolumn]{revtex4}
\usepackage{amsmath,amssymb}

\newcommand{\eqa}{\begin{eqnarray}}
\newcommand{\neqa}{\end{eqnarray}}
\newcommand{\be}{\begin{equation}}
\newcommand{\ee}{\end{equation}}

\renewcommand{\texttt}{{}}
\usepackage{graphicx}

\begin{document}

\title{%
Space-Time Structure of Loop Quantum Black Hole}

\author{Leonardo Modesto}

\affiliation{ Perimeter Institute for Theoretical Physics, 31 Caroline St., Waterloo, ON N2L 2Y5, Canada}

\date{\small\today}

\begin{abstract} \noindent
In this paper we have improved
the semiclassical analysis of loop quantum black hole
(LQBH) 
in the conservative approach of constant polymeric
parameter. 
In particular we have focused our attention on the space-time structure.
We have introduced a very simple 
modification of the spherically 
symmetric Hamiltonian constraint in 
its holonomic version. 
The new quantum constraint reduces to the classical constraint 
when the polymeric parameter $\delta$ goes to zero.
Using this modification 
we have obtained a large class 
of semiclassical solutions parametrized by
a generic function $\sigma(\delta)$.
We have found that 
only a particular choice of this function
reproduces the black hole solution
with the correct asymptotic flat limit.
In $r=0$ the semiclassical metric is regular
and the Kretschmann invariant
has a maximum  peaked in $r_{\rm max} \sim l_P$.
The radial position of the pick  
does not depend on the
black hole mass and the polymeric parameter $\delta$.
The semiclassical solution is very similar to the 
Reissner-Nordstr\"om metric. We have constructed 
the Carter-Penrose diagrams explicitly,  
giving a causal description of the space-time and its 
maximal extension. 
The LQBH metric interpolates 
between two asymptotically $flat$ regions, the $r\rightarrow \infty$ 
region and the $r\rightarrow 0$ region.
We have studied the thermodynamics of the semiclassical solution. 
The temperature, entropy and the evaporation process 
are regular and could be defined independently from
the polymeric parameter $\delta$.
We have studied the particular metric when 
the polymeric parameter goes towards to zero. This metric is regular in $r=0$
and has only one event horizon in $r = 2m$. The Kretschmann  invariant
maximum depends only on $l_P$. The polymeric parameter $\delta$
does not play any role in the black hole singularity resolution.
The thermodynamics is the same.


\end{abstract}

\maketitle

\section*{INTRODUCTION}

Quantum gravity is the theory  attempting to reconcile general relativity and quantum
mechanics. 
In general relativity the space-time is dynamical, then 
it is not possible to study other interactions on a fixed background because
the background itself is a dynamical field.
 The theory called
 ``loop quantum gravity" (LQG) \cite{book}
is the most widespread nowadays.
This is one of the non perturbative and 
background independent approaches to quantum gravity.
LQG is a quantum geometric 
fundamental theory that reconciles general relativity 
and quantum mechanics at the Planck scale
and we expect that this theory could resolve the
classical singularity problems of General Relativity.
Much progress has been done in this direction in the 
last years. 
In particular, the application of LQG technology 
to early universe in the context of minisuperspace models
have solved the initial singularity problem \cite{Boj}, \cite{MAT}.

Black holes 
are another interesting place for testing the validity of LQG.
In the past years 
applications of LQG ideas to the Kantowski-Sachs space-time 
\cite{KS} 
lead to some interesting results in this field. 
In particular, it has been 
showed 
\cite{work1} \cite{work2} that it is 
possible to solve the black hole singularity problem by using 
tools and ideas developed in full LQG. 
Other remarkable results have been obtained in the non homogeneous 
case \cite{GP}.

There are also works of semiclassical nature which try to solve the black hole singularity 
problem \cite{SS},\cite{SS2}, \cite{SS2}.
 In these papers the authors use an effective Hamiltonian constraint 
 obtained replacing the Ashtekar connection $A$ with the holonomy $h(A)$
 and they solve the classical Hamilton equations of motion exactly 
 or numerically.
 In this paper we try to improve the semiclassical analysis 
 introducing a very simple modification to the holonomic version of the
 Hamiltonian constraint.
 The main result is that
 the minimum area \cite{LoopOld} of full LQG is the
 fundamental ingredient to solve
 the black hole space-time singularity problem in $r=0$.
 The $S^2$ sphere bounces on the minimum area $a_0$ of LQG 
 and the singularity disappears. 
 We show the Kretschmann  invariant is regular in all space-time and
 the position of the maximum is independent on mass and on polymeric parameter
 introduced to define the holonomic version of the scalar constraint.
 The radial position of the curvature maximum depends only on $G_N$ and $\hbar$.
 
This paper is organized as follows.  In the first section we 
recall the classical Schwarzschild solution in Ashtekar's variables and we introduce
a class of Hamiltonian constraints expressed in terms of holonomies 
that reduce to the classical one in the limit where the polymer parameter $\delta \rightarrow 0$.
We solve the Hamilton equations of motion obtaining the semiclassical black 
hole solution for a particular choice of the quantum constraint. 
In the third section we show the regularity of the solution by studying 
the Kretschmann  operator and we write the solution in a very simple form 
similar to the Reissner-Nordstr\"om solution for a black hole with
mass and charge. In section four we study the space-time structure 
and we construct the Carter-Penrose diagrams. In section five section  we
show the solution has a Schwarzschild core in $r\sim 0$. 
In section six we analyze the black hole thermodynamic
calculating temperature, entropy and evaporation. In section seven 
we calculate the limit $\delta\rightarrow 0$ of the metric and we obtain
 a regular semiclassical solution with the same thermodynamic 
properties but with only one event horizon at the Schwarzschild radius.
We analyze the causal space-time 
structure and construct the Carter-Penrose diagrams.



\section{Schwarzschild solution in Ashtekar variables}
In this section we recall the classical Schwarzschild solution inside the event 
horizon \cite{work1} \cite{work2}.
For the homogeneous but non isotropic Kantowski-Sachs
space-time the Ashtekar's variables \cite{variables} are 
\begin{eqnarray}
&& A= \tilde{c} \tau_3 d x + \tilde{b} \tau_2 d \theta - \tilde{b} \tau_1 \sin \theta d \phi + \tau_3 \cos \theta d \phi,
\nonumber \\
&&E = \tilde{p}_c \tau_3 \sin \theta \frac{\partial}{\partial x} + \tilde{p}_b \tau_2 \sin \theta \frac{\partial}{\partial \theta} - \tilde{p}_b \tau_1 \frac{\partial}{\partial \phi}.
\label{contriad}
\end{eqnarray}
The components variables in the phase space have length dimension 
$[\tilde{c}]=L^{-1}$, $[\tilde{p}_c]=L^{2}$, $[\tilde{b}]=L^{0}$, $[\tilde{p}_b]=L$.
The Hamiltonian constraint is 
\begin{eqnarray}
&& \hspace{-0.8cm} 
\mathcal{C}_{H} = - \int \frac{N d x \sin \theta d \theta d \phi}{8 \pi  G_N \gamma^2} \nonumber \\
&& \hspace{2cm} 
\left[ (\tilde{b}^2 + \gamma^2) \frac{\tilde{p}_b \, {\rm sgn} (\tilde{p}_c)}{\sqrt{|\tilde{p}_c|}} + 2 \tilde{b} \tilde{c} \, \sqrt{|\tilde{p}_c|} \right]. 
\label{Ham1}
\end{eqnarray}
Using the general relation $E^a_i E^b_j \delta^{i j}= {\rm det}(q) q^{ab}$ ($q_{ab}$ is the metric on 
the spatial section) we obtain 
$q_{ab} = (\tilde{p}_b^2/|\tilde{p}_c|, |\tilde{p}_c|, |\tilde{p}_c| \sin^2 \theta)$.

We restrict integration over $x$ to a finite interval $L_0$ and the Hamiltonian 
takes the form \cite{work2}
\begin{eqnarray}
\mathcal{C}_{H} = - \frac{N}{2 G_N \gamma^2} \left[ (b^2 + \gamma^2) \frac{p_b \, {\rm sgn} (p_c)}{\sqrt{|p_c|}} + 2 b c \, \sqrt{|p_c|} \right].
\label{Ham2}
\end{eqnarray}
The rescaled variables are: $b=\tilde{b}$, $c= L_0 \tilde{c}$, $p_b = L_0 \tilde{p}_b$, $p_c=\tilde{p}_c$. The length dimensions of the new phase space variables are: 
$[c]=L^{0}$, $[p_c]=L^{2}$, $[b]=L^{0}$, $[p_b]=L^2$. 
From the symmetric reduced connection and density 
triad we can read the components variables in the phase space:
$(b, p_b)$, $(c, p_c)$, 
with Poisson algebra $\{c, p_c \} = 2 \gamma G_N$, $\{b, p_b \} = \gamma G_N$.
We choose the gauge $N=\gamma \, \sqrt{|p_c|} \, {\rm sgn} (p_c)/ b$ and the
Hamiltonian constraint reduce to 
\begin{eqnarray}
\mathcal{C}_{H} = - \frac{1}{2 G_N \gamma} \left[ (b^2 + \gamma^2) p_b/b + 2 c p_c \right].
\label{Ham3}
\end{eqnarray}
The 
Hamilton equations of motion are
\begin{eqnarray}
&& \dot{b} = \{ b, \mathcal{C}_{H} \} = - \frac{b^2 + \gamma^2}{2 b}, \nonumber \\
&& \dot{p_b} = \{ p_b, \mathcal{C}_{H} \} =  \frac{1}{2} \Big[p_b -\frac{\gamma^2 p_b}{b^2}\Big],
\nonumber \\
&&  \dot{c} = \{ c, \mathcal{C}_{H} \} = - 2 c, \nonumber \\ 
&&\dot{p_c} = \{ p_c, \mathcal{C}_{H} \} = 2 p_c.
\label{Eq.1}
\end{eqnarray}
The solutions of equations (\ref{Eq.1}) using the time parameter $t \equiv e^T$ 
and redefining the integration constant $\equiv e^{T_0} = 2 m$ 
(see the papers in \cite{work1} \cite{work2}) are 
\begin{eqnarray}
&& b(t) = \pm \gamma \sqrt{2m/t - 1}, \nonumber \\
&& p_b(t) = p_b^{0} \sqrt{t(2 m - t)} \nonumber \\
 && c(t) = \mp \gamma m p_b^{0} t^{-2}, \nonumber \\
 && p_c(t) = \pm t^2.
 \label{Sol.1}
\end{eqnarray}
This is exactly the Schwarzschild solution inside and also 
outside the event horizon as we
can verify passing to the metric form defined by 
$h_{ab} = \mbox{diag}(p_b^2/|p_c| L_0^2, |p_c|, |p_c|  \sin^2 \theta)$ ($m$ contains the gravitational constant parameter $G_N$).
The line element is 
\begin{eqnarray}
ds^2=-N^2 \frac{dt^2}{t^2} + \frac{p_b^2}{|p_c| \, L_0^2} dx^2 + |p_c| (\sin^2 \theta d \phi^2+d\theta^2).
\label{line-element}
\end{eqnarray}
Introducing the solution (\ref{Sol.1}) in (\ref{line-element}) 
we obtain the Schwarzschild solution in all space-time except in $t=0$ where 
the classical curvature singularity is localized and except in $r= 2m$ where there is 
a coordinate singularity 
 \begin{eqnarray}
ds^2=- \frac{dt^2}{\frac{2 m}{t} -1} + \frac{(p_b^{0})^{2}}{L_0^2} \left(\frac{2 m}{t} -1\right) dx^2 + t^2 
d \Omega^{(2)},
\label{line-element2}
\end{eqnarray}
where $d \Omega^{(2)} = \sin^2 \theta d \phi^2+d\theta^2$.
To obtain the Schwarzschild metric we choose $L_0 = p_b^{0}$.  
In this way we fix the radial cell to have length $L_0$ 
and $p_b^{0}$ disappears from the metric. 
In the semiclassical LQBH metric $p_b^{0}$ does not disappears
fixing $L_0$.
At this level we have not fixed 
$p_b^{0}$ but only the dimension of the radial cell. 
This is the correct choice to
reproduce the Schwarzschild solution. 
We have defined the dimension of the
cell in the $x$ direction to be $L_0 = p_b^{0}$ obtaining the correct Schwarzschild 
metric in all space time, we will do the same choice for the semiclassical metric.
With this choice 
$p_b^{0}$ will not disappears from the semiclassical metric and
in particular from the $p_c(t)$ solution.
We will use the minimum area of the full theory to fix $p_b^{0}$.
For the semiclassical solution at the end of section (\ref{core}) we will give also a possible 
physical interpretation of $p_b^{0}$.

\section{A general class of Hamiltonian constrains
}
The correct dynamics of loop quantum gravity is the main problem
of the theory. LQG is well defined at kinematical level but it is not        
clear what is the correct version of the Hamiltonian constraint, or
more generically,
in the covariant approach, what is the correct spin-foam model
\cite{SpinFoams}.
An empirical 
principle to construct the correct Hamiltonian constraint
is to recall the correct semiclassical limit \cite{SemiLim}.
When we impose spherical symmetry and homogeneity, 
the connection and density triad assume 
the particular form given in (\ref{contriad}).
We can choose a large 
class of Hamiltonian constraints, expressed in terms of holonomies $h^{(\delta)}(A)$, 
which reduce to the same classical one (\ref{Ham3}) 
when the polymeric parameter $\delta$ goes towards to zero.
We introduce
a parametric function $\sigma(\delta)$
that labels the elements in the class of 
Hamiltonian constraints compatible with 
spherical symmetry and homogeneity.
We call ${\mathcal C}_{LQG}$ the constrain for the full theory 
and ${\mathcal C}_{\sigma(\delta)}$ the constraint for the homogeneous 
spherical minisuperspace model. 
The reduction from the full theory to the minisuperspace model
is 
\begin{eqnarray}
&& \mathcal{C}_{LQG} \rightarrow \mathcal{C}_{\sigma(\delta)},
\end{eqnarray}
where the arrow represents the spherical symmetric reduction 
of the full loop quantum gravity hamiltonian constraint. 
To obtain the classical Hamiltonian constraint (\ref{Ham3}) in the limit 
$\delta \rightarrow 0$ we recall that the function $\sigma(\delta)$ satisfies 
the following condition
\begin{eqnarray}
\lim_{\delta \rightarrow 0} \, \sigma(\delta) = 1 \,\,\, \rightarrow \,\,\, \lim_{\delta \rightarrow 0} \mathcal{C}_{\sigma(\delta)} = \mathcal{C}_H.
\end{eqnarray}
We are going to show that just one particular choice of $\sigma(\delta)$ gives the 
correct asymptotic flat limit for the Schwarzschild black hole.
In fact the asymptotic
$boundary$ $condition$ selects the particular form of the function $\sigma(\delta)$.

The classical Hamiltonian constraint can be written in the following form
\begin{eqnarray}
 \mathcal{C}_{H} = \frac{1}{\gamma^2} \int d^3 x \epsilon_{i j k} e^{-1} E^{a i} E^{b j} 
\left[\gamma^2 \Omega^k_{a b} - ^0 F^k_{a b} \right], 
\end{eqnarray}
where $\Omega=- \sin \theta \tau_3 d \theta \wedge d \phi$ and 
$^0 F = dK +[K,K]$ ($K$ is the extrinsic curvature, $A=\Gamma + \gamma K$
and $\Gamma = \cos \theta \, \tau_3 \, d \phi$
).
The holonomies in the directions $x, \theta, \phi$ for a generic path $\ell$ are 
defined by
\begin{eqnarray}
&&  h_1^{(\ell)} = \cos \frac{\ell c}{2} + 2 \tau_3 \sin \frac{\ell c}{2}, \nonumber \\
&&  h_2^{(\ell)} = \cos \frac{\ell b}{2} - 2 \tau_1 \sin \frac{\ell b}{2}, \nonumber \\
&& h_3^{(\ell)} = \cos \frac{\ell b}{2} + 2 \tau_2 \sin \frac{\ell b}{2}. 
\end{eqnarray}
We define 
the field straight $^0F^{i}_{a b}$ in terms of holonomies in the following way
\begin{eqnarray}
&&\hspace{-0.2cm}^0F^{i}_{a b} \tau_i = ^0F^{i}_{i j} \,^0 \omega^i_a \, ^0 \omega^j_b \,
\left( \frac{h_i^{(\delta_i)} h_j^{(\delta_j)} h_i^{(\delta_i) -1} h_j^{( \delta_j )-1}}{\delta^2} \right), \nonumber \\ 
&& \delta_i = (\delta c, \sigma(\delta) \delta b, \sigma(\delta) \delta b),
\label{FS}
\end{eqnarray}
it's a simple exercise to verify that when $\delta \rightarrow 0$ 
(\ref{FS}) we obtain the classical field straight.
The Hamiltonian constraint in terms of holonomies is 
\begin{eqnarray}
&& \hspace{0.0cm} \mathcal{C}_{\sigma(\delta)} = 
 \frac{ - N}{(8 \pi G_N)^2 \gamma^3 \delta^3} \, \times \nonumber \\
&& \times \, {\rm Tr} \Big[\sum_{ijk} \epsilon^{ijk} h_i^{(\delta_i)} h_j^{(\delta_j)} h_i^{(\delta_i) -1}
h_j^{(\delta_j) -1}
 h_k^{(\delta)}
\left\{h_k^{(\delta) -1}, V \right\} \nonumber \\
&& \hspace{3.8cm}+ 2 \gamma^2 \delta^2 \tau_3 h_1^{(\delta)}\left\{h_1^{(\delta) -1}, V\right\}     
    \hspace{-0.0cm} \Big]  \nonumber \\
&& \hspace{0.85cm} = - \frac{ N}{2 G_N \gamma^2 } 
\Bigg\{ 2 \frac{\sin \delta c}{\delta} \ \frac{\sin (\sigma(\delta) \delta b)}{\delta} \ \sqrt{|p_c|} \nonumber \\
&& \hspace{1.5cm}+ \left(\frac{\sin^2 (\sigma(\delta) \delta b)}{\delta^2} + \gamma^2  \right) \frac{p_b \ \mbox{sgn}(p_c)}{\sqrt{|p_c|}} \Bigg\}.
\label{CH}
\end{eqnarray} 
$V = 4 \pi \sqrt{|p_c|} p_b$ is the spatial section volume.
We have introduced modifications depending on the function $\sigma(\delta)$ 
only in the field straight but this is sufficient to have 
a large class of semiclassical hamiltonian constraints compatible 
with spherical simmetry.
The Hamiltonian constraint $\mathcal{C}^{\delta}$ in (\ref{CH})
can be substantially simplified   
in the gauge 
$
N = (\gamma \sqrt{|p_c|} \mbox{sgn}(p_c) \delta)/(\sin \sigma(\delta) \delta b)$
\begin{eqnarray}
&& \mathcal{C}_{\sigma(\delta)} = - \frac{1}{2 \gamma G_N} 
\Big\{ 2 \frac{\sin \delta c}{\delta}  \  p_c + \nonumber \\
&& \hspace{1.5cm}\left(\frac{\sin \sigma(\delta)\delta b}{\delta} + 
\frac{\gamma^2 \delta }{\sin \sigma(\delta) \delta b} \right) 
p_b \Big\}.
\label{FixN}
\end{eqnarray}
From (\ref{FixN}) we obtain two independent sets of equations of motion on the 
phase space 
\begin{eqnarray}
&& \dot{c} = - 2 \frac{\sin \delta c}{\delta}, \nonumber \\
&& \dot{p_c} = 2   p_c \cos \delta c,
\nonumber \\
&&\dot{b} = - \frac{1}{2} \left(\frac{\sin \sigma(\delta) \delta b}{\delta} + \frac{ \gamma^2 \delta}{\sin \sigma(\delta) \delta b} \right), \nonumber \\ 
&& \dot{p_b} =  \frac{\sigma(\delta)}{2} \, \cos \sigma(\delta)\delta b \left(1 - \frac{ \gamma^2 \delta^2}{\sin^2 \sigma(\delta) \delta b} \right) p_b.
\end{eqnarray}  
Solving the first three equations and using the Hamiltonian constraint $\mathcal{C}^{\delta} =0$,
with the time parametrization $e^T = t$ and imposing to have the 
Schwarzschild event horizon in $t=2 m$, we obtain
\begin{eqnarray}
 && c(t) =  \frac{2}{\delta} \arctan \Big( \mp \frac{\gamma \delta m p_b^{0}}{2 t^2}  \Big),
\nonumber \\
&& p_c (t) = \pm 
\frac{1}{t^2} 
 \Big[\Big(\frac{\gamma \delta m p_b^{0}}{2}\Big)^2  + t^4 \Big], \nonumber \\
&& \cos \sigma(\delta) \delta b = \nonumber \\
&& \hspace{0.5cm} = \rho(\delta)
\left[ \frac{ 1 - \Big(\frac{ 2 m}{t} \Big)^{\sigma(\delta) \rho(\delta)}
\mathcal{P}(\delta) }
{ 1 + \Big(\frac{ 2 m}{t} \Big)^{\sigma(\delta) \rho(\delta)}
\mathcal{P}(\delta)}
\right], \nonumber \\
&& p_b(t) = -  \frac{2 \ \sin \delta c \
\sin \sigma(\delta) \delta b \ p_c }{\sin^2 \sigma(\delta) \delta b + \gamma^2 \delta^2},
\label{Sol.cpcbpb}
\end{eqnarray}
where we have defined the quantities 
\begin{eqnarray}
&& \rho(\delta) = \sqrt{1 + \gamma^2 \delta^2}, \nonumber \\
&& \mathcal{P}(\delta) = \frac{\sqrt{1 + \gamma^2 \delta^2} -1}{\sqrt{1 + \gamma^2 \delta^2} +1}.
\label{oro}
\end{eqnarray}
Now we focus our attention on the term $(2m /t)^{\sigma(\delta) \rho(\delta)}$.
The choice of this term and in particular the choice of the exponent will be crucial to have the 
correct flat asymptotic limit. 
The exponent is in the form $(2m/t)^{1+\epsilon}$ and expanding in powers of
the small parameter $\epsilon \sim \delta^2$ we obtain $(2m/t)^{1+\epsilon} \sim - (2m/t) \log(t/2m)$ 
at large distance ($t\gg 2m$) (we remember that outside the event horizon the coordinate
t plays the rule of spatial radial coordinate). It is straightforward to see that exists only one 
possible way to obtain the correct asymptotic limit and it is given by the choice $\sigma(\delta)=1/\sqrt{1+\gamma^2\delta^2}$. In other words we can say that any function $x^{\epsilon} \sim \epsilon \log(x)$
diverges logarithmically for small $\epsilon$ and large distance ($x\gg1$).

Let as take 
$\sigma(\delta) = 1/\sqrt{1 + \gamma^2 \delta^2}$. In force of
the correct large distance limit and in force also of the regularity of the curvature invariant 
in all space time, we will extend the solution outside the event horizons with the
redefinition $t \leftrightarrow r$. I will come back to this extension in the next section.

A crucial difference with the classical Schwarzschild solution is that 
$p_c$ has a minimum in $t_{min} = (\gamma \delta m p_b^{0}/2)^{1/2}$,
and $p_{c}(t_{min}) = \gamma \delta m p_b^{0}$.
The solution has a spacetime structure very similar to the
Reissner-Nordstr\"om metric and presents an inner horizon in 
\begin{eqnarray}
r_{-} = 2 m {\mathcal P}(\delta)^2 = 
2 m \left(\frac{2+\gamma^2 \delta^2-2\sqrt{1+\gamma^2 \delta^2}}{2+\gamma^2 \delta^2+2\sqrt{1+\gamma^2 \delta^2}}\right).
\label{rmeno}
\end{eqnarray}
For $\delta \rightarrow 0$, $r_{-} \sim m \gamma^4 \delta^4/8$.
We observe that the inside horizon position $r_-\neq 2m \,\, \forall \gamma \in\mathbb{R}$
(we recall $\gamma$ is the Barbero-Immirzi parameter). 
\begin{figure}
 \begin{center}
 \includegraphics[height=6cm]{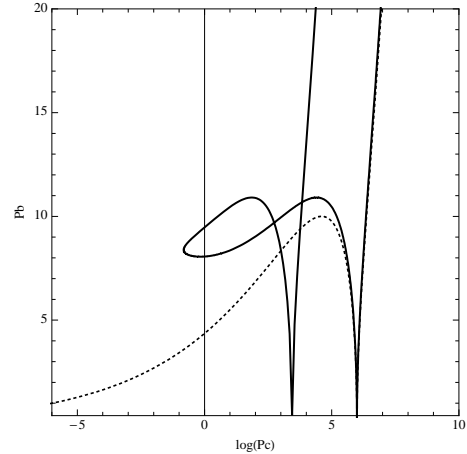}
  \hspace{1cm}
  \end{center}
  \caption{  \label{pbpc} 
   Semiclassical dynamical trajectory on the plane $(\sqrt{|p_b^2|}/p_b^{0}, \log(p_c))$ 
    for positive values of $p_c$.  
  The dashed trajectory corresponds to
                 the classical Schwarzschild solution and the 
                 continuum trajectory corresponds to the semiclassical solution. 
                  The plot refers to $m = 10$, $p_b=1/10$ and  $\gamma \delta = \log(4)/\pi$.   }
  \end{figure}
Now we study the trajectory in
the plane $(p_b/p_b^{0}, \log(p_c))$ and we compare the result with
the Schwarzschild solution.
In Fig.\ref{pbpc}  we have a parametric plot of $(|p_b|, \log(p_c))$;
we can follow the trajectory from $t > 2m$ where 
the classical (dashed trajectory) and the semiclassical (continuum trajectory) 
solution are very close. For $t = 2m$,
$p_c \rightarrow (2m)^2$ and $p_b \rightarrow 0$ (this point 
corresponds to the Schwarzschild radius).
From this point decreasing $t$ we reach 
a minimum value for $p_{c,m} \equiv p_c(t_{min}) >0$. From $t=t_{ min}$, $p_c$
starts to grow again until $p_b=0$, this point corresponds to a new horizon 
in $t =r_{-}$ localized. In the time interval $t < t_{min}$, $p_c$ grows
together with $|p_b|$ and as it is very clear from the 
picture the solution approach the second specular black hole for
$t\rightarrow 0$. In particular we have a second $flat$ asymptotic  region
for $t \sim 0$.

\subsection*{Metric form of the solution.}

In this section we write 
the solution in  
the metric form and we extend that to the all space-time.
We recall the Kantowski-Sachs metric is 
$ds^2 = - N^2(t) dt^2 + X^2(t) dx^2 + Y^2(t)(d \theta^2 + \sin \theta d \phi^2)$.
The metric components are related to the connection variables by 
\begin{eqnarray}
&& N^2(t) = \frac{\gamma^2 \delta^2 |p_c(t)|}{ t^2 \sin^2 \sigma(\delta)\delta b}, \,\,\,\,\,\,\, X^2(t) = \frac{p_b^2(t)}{L_0^2 |p_c(t)|} \Omega(\delta), \nonumber\\
&& \, Y^2(t) =|p_c(t)|. 
\label{metric}
\end{eqnarray}
We have introduced $\Omega(\delta)$ by a coordinate transformation 
$x\rightarrow \sqrt{\Omega(\delta)} \, x$,
\begin{eqnarray}
 \Omega(\delta) = 16 (1 + \gamma^2 \delta^2)^2/(1 + 
     \sqrt{1 + \gamma^2 \delta^2})^4
\label{omega}
\end{eqnarray}
 This coordinate transformation 
is useful to obtain the Minkowski metric in the limit $t\rightarrow \infty$.
The explicit form of the lapse function $N(t)^2$ in terms of the coordinate $t$ is 
\begin{eqnarray}
N^2(t) = \frac{\gamma^2 \delta^2 \Big[ \Big(\frac{\gamma \delta m p_b^{0}}{2 t^2} \Big)^2 +1 \Big]}{
  1 - \rho^2( \delta)
\left[ \frac{1 - \left(\frac{2m}{t} \right)\mathcal{P(\delta)}}
{{1 + \left(\frac{2m}{t} \right)
\mathcal{P(\delta)}}}
\right]^2   
}.
\label{N2metric}
\end{eqnarray}
Using the second relation in (\ref{metric}) we can obtain the $X^2(t)$ metric component, 
\begin{eqnarray}
&& \hspace{-0.6cm}X^2(t) 
= 
\frac{(2 \gamma \delta m )^2   \Omega(\delta)
                         \left(1 - \rho^2(\delta)
\left[ \frac{ 1 - \frac{2m}{t} 
\mathcal{P}(\delta)}
{1 + \frac{2m}{t} 
\mathcal{P}(\delta) }
\right]^2   \right) \  t^2 }{
\rho^4(\delta) \left(1 -
\left[ \frac{1 - \frac{2m}{t} \mathcal{P}(\delta)}
{1 + \frac{2m}{t} \mathcal{P}(\delta)}
\right]^2\right)^2
\Big[\Big(\frac{\gamma \delta m p_{b}^{0}}{2}\Big)^2  + t^4 \Big]
}.\nonumber \\
&&
\label{X2}
\end{eqnarray}
The function $Y^2(t)$ corresponds to $|p_c(t)|$ given in (\ref{Sol.cpcbpb}).
The metric obtained has the correct asymptotic limit for $t\rightarrow + \infty$ and 
in fact  $N^2(t \rightarrow 0) \rightarrow - 1$, 
$X^2(t \rightarrow 0) \rightarrow - 1$, 
$Y^2(t \rightarrow 0) \rightarrow t^2$.
The semiclassical metric goes to a flat limit also 
for $t\rightarrow 0$. We can say that LQBH interpolates 
between two asymptotic flat region of the space-time.
The metric obtained in this paper has the correct flat asymptotic 
limit for $t\rightarrow +\infty$ and reproduce the Minkowski 
metric for $m\rightarrow 0$. Both those limit are not  satisfied 
in the work \cite{SS}. The small modification 
introduced in the holonomy form of the Hamiltonian is necessary
for those two fundamental consistency limit.

\section{LQBH in all space-time} \label{LQBH}

\begin{figure}
 \begin{center}
  \includegraphics[height=6cm]{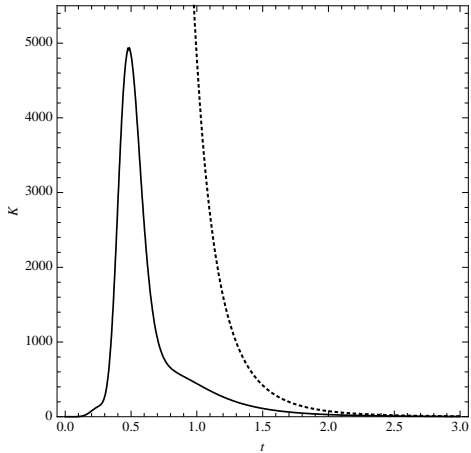}
    \end{center}
  \caption{\label{K0} 
 Plot of the
 Kretschmann  scalar 
  invariant ${\rm R}_{\mu \nu \rho \sigma} {\rm R}^{\mu \nu \rho \sigma}$ 
for $m = 10$, $p_{b}^{0} =1/10$ and $\gamma \delta \sim 1$, $\forall t \geqslant 0$; the large 
 $t$ behaviour 
  is $1/t^6$.}
  \end{figure}

In this section we extend the (metric) semiclassical solution obtained 
obtained in the previous section to all space-time.
As explained in the previous subsection the metric solution has the correct flat limit
for $t\rightarrow 0$ and goes to Minkowski for $m \rightarrow 0$.
Now we shaw that the Kretschmann  scalar 
$K= \rm R_{\mu \nu \rho \sigma} \rm R^{\mu \nu \rho \sigma}$
is regular in all space-time.
In terms of $N(t)$, $X(t)$ and $Y(t)$ the Kretschmann  scalar is 
\begin{eqnarray}
&& \hspace{-0.0cm}{\rm R_{\mu \nu \rho \sigma} \rm R^{\mu \nu \rho \sigma}} = \nonumber \\
&& =  \hspace{-0.1cm} 4 \Bigg[
\left(\frac{1}{ X N} \frac{d}{d t} \left( \frac{1}{N} \frac{d X}{dt} \right)\right)^2 + 
2 \left(\frac{1}{ Y N} \frac{d}{d t} \left( \frac{1}{N} \frac{d Y}{dt} \right)\right)^2  \nonumber \\
&& \hspace{-0.1cm}+ 2\left(   \frac{1}{ X N} \frac{d X}{d t} \, \frac{1}{ Y N} \frac{d Y}{d t} \right)^2 +
\frac{1}{Y^4 N^4}\left(N^2 + \Big(\frac{d Y}{d t}\Big)^2 \right)^2 
\hspace{-0.1cm}
\Bigg] \hspace{-0.08cm} . \nonumber \\
&&
\nonumber \\
&&
\label{CUR}
\end{eqnarray}
In Fig.\ref{K0} is plotted a graph of $K$, it is regular in all space-time and the large
$t$ behavior is the classical singular scalar
${\rm R}_{\mu \nu \rho \sigma} {\rm R}^{\mu \nu \rho \sigma} = 48 m^2/t^6$.
  
\begin{figure}
 \begin{center}
  \includegraphics[height=6cm]{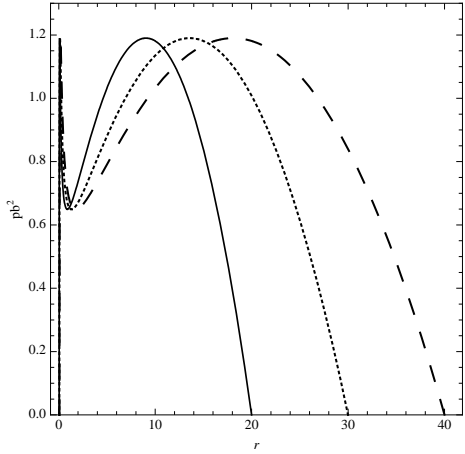}
    \end{center}
  \caption{\label{pb3} 
 Plot of $p_b^2(t)$ for different values of the mass ($m=10, 15, 20$).
 Max (absolute) and Min (relative) of $p_b^2(t)$ are independent of the mass $m$.}
  \end{figure}

  What about $p_b^{0}$?
  Now we fix the parameter $p_b^{0}$ using the full theory (LQG).
  \noindent In particular we choose $p_b^{0}$ in such way the position $r_{\rm Max}$ 
  of the Kretschmann  invariant maximum is independent of the black hole mass. 
  This means  the $S^2$ sphere bounces on a minimum radius 
  that is independent from the mass of the black hole and from $p_b^{0}$ 
  and depends only on $l_P$.
  We consider the solution $p_c(t)$ and we impose the
  minimum area $A_{\rm Min}= 4 \pi \gamma \delta m p_b^{0}$ 
  of the $S^2$ sphere to be equal to the minimum gap area of 
  loop quantum gravity $a_0 = 2 \sqrt{3} \pi \gamma  l_P^2$. 
  With the choice $\gamma \delta  m p_b^{0} = a_0/4 \pi$
  we obtain a significative physical result. 
  We have {\em not} impose $p_c(t)$ to have a minimum in $a_0$ but
  we have just impose that the minimum of $p_c(t)$ is 
  the minimum area of the full theory. The minimum area of the two sphere
  is a result and not a request.  
  We observe that this choice of $p_b^{0}$ fixes the absolute maximum and relative minimum 
  of $p_b(t)$ to be independent of the mass $m$ as this is manifest from the plot 
  in Fig.\ref{pb3}.
  
  We want to provide an argument to support the choice $p_b^{0} \sim a_0/m$.
  In the paper \cite{h} it is shown the phase space is parametrized by $m$ and 
  the conjugated momentum $p_m$ and it is shown that are both 
  constants of motion (in our notation $p_m=p_b^{0}$). 
  As usual in elementary quantum mechanics to derive the 
  Heisenberg uncertainty relation, we can introduce 
  the state $| \phi \rangle  = (\hat{m} + i \lambda \hat{p}_m)|\psi \rangle$,
  where $\hat{m}$ and $\hat{p}_m$ are the mass and momentum operators 
  and $\lambda \in \mathbb{R}$.
  From the positive norm 
  $\langle \phi | \phi \rangle =\langle \hat{m}^2  \rangle +i \lambda \langle [\hat{m}, \hat{p}_m] \rangle
  +\lambda^2 \langle \hat{p}^2_m \rangle\geqslant 0$ we have the discriminant, of second order 
  in $\lambda$, is negative or zero. The condition on the discriminant gives 
  $\langle \hat{m}^2  \rangle \langle \hat{p}_m^2  \rangle \geqslant -
   \langle [\hat{m}, \hat{p}_m]  \rangle^2/4$. 
   Introducing the commutator $[\hat{m}, \hat{p}_m] = i l_P^2$ 
   we obtain $\langle \hat{m}^2  \rangle \langle \hat{p}_m^2  \rangle \geqslant l_P^4$.
  We can calculate $\langle \hat{m}^2 \rangle$
  on semiclassical gaussian states, 
  \begin{eqnarray}
 \Psi(m)_{m_0, p_0} = \frac{{\rm e}^{-\frac{(m-m_0)^2}{4\Delta^2}} {\rm e}^{\frac{i p_0 m}{l_P^2}}}{(2 \pi \Delta^2)^{1/4}},
 \end{eqnarray}
 and the result is $\langle \hat{m}^2 \rangle =  4 m_0^2$ (for $\Delta =\sqrt{3} m_0$).
  Using the Heisenberg uncertainty relation
  we determine $ \langle \hat{p}_m^2 \rangle= l_P^4/16 m_0^2$.
  If we identify $\langle \hat{p}_m^2 \rangle=(p_b^0)^{2}$ we obtain 
  $m_0 p_b^{0}= l_P^2/4$,
  which is exactly $m p_b^{0} = a_0/4 \pi \gamma \delta$ for $\delta = 2 \sqrt{3}$,
  $a_0=2 \sqrt{3} \pi \gamma  l_P^2$ and $m_0\equiv m$. 
 We have introduced explicitly all the coefficients but the main result is $p_b^{0} \sim a_0/m$.
 However the presented here is just an argument and not a proof.

  At the end of section (\ref{core}) we will give a physical interpretation of 
  $p_b^{0}$.  
  
 We now want underline the similarity between the equation of motion for
  $p_c(t)$ and the Friedmann equation of loop quantum cosmology.
  We can write the differential equation for  $p_c(t)$ in the following form
  \begin{eqnarray}
  \left(\frac{\dot{p}_c}{p_c} \right)^2= 4  \left(1- \frac{a_0^2}{16 \pi^2 p_c^2}\right).
   \end{eqnarray}
  From this equation is manifest that $p_c$ bounces on the value $a_0/4 \pi$.
  This is quite similar to the loop quantum cosmology bounce \cite{AshtekarSingh}.
 
\begin{figure}
 \begin{center}
 \includegraphics[height=5cm]{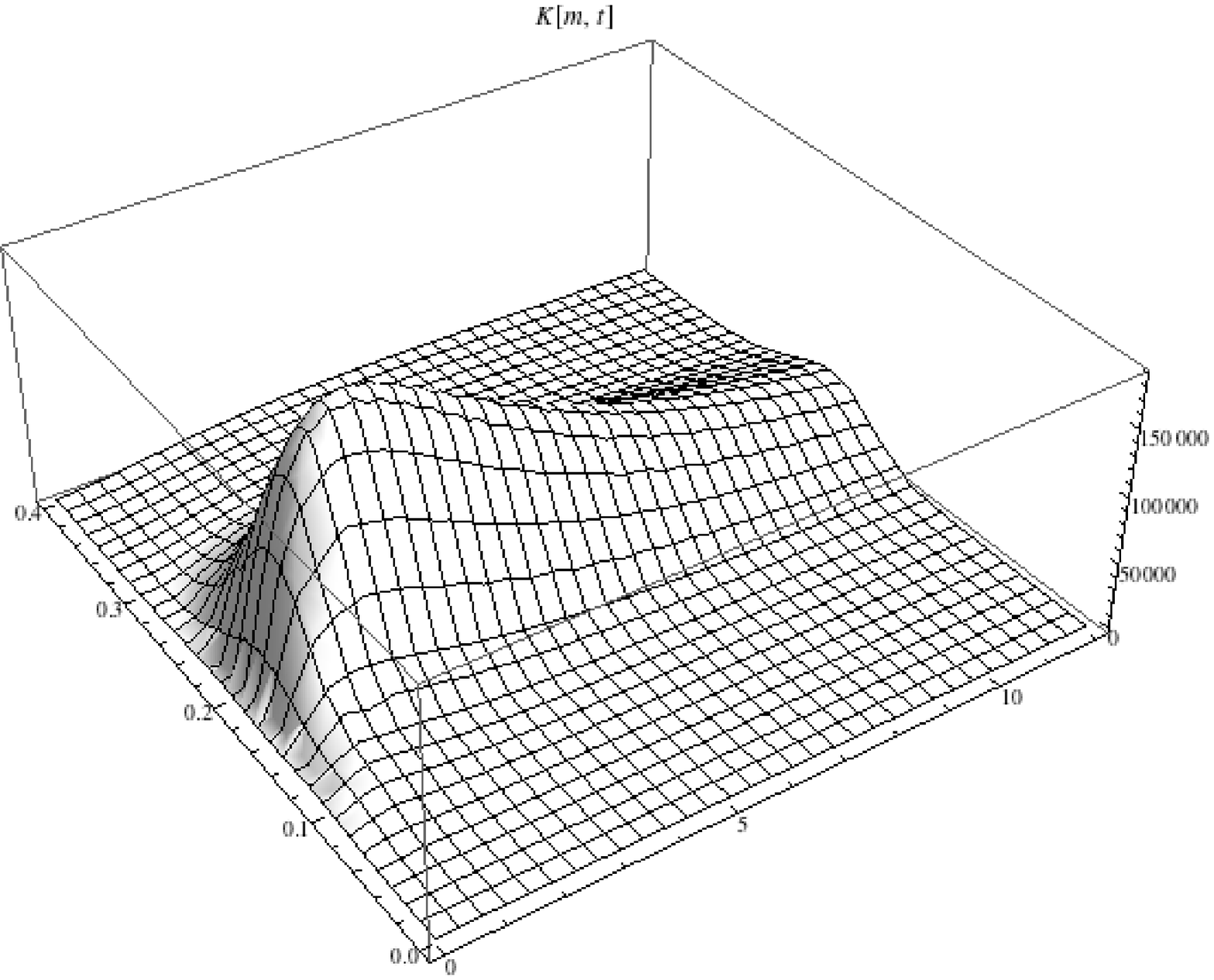}\\ 
  \includegraphics[height=5cm]{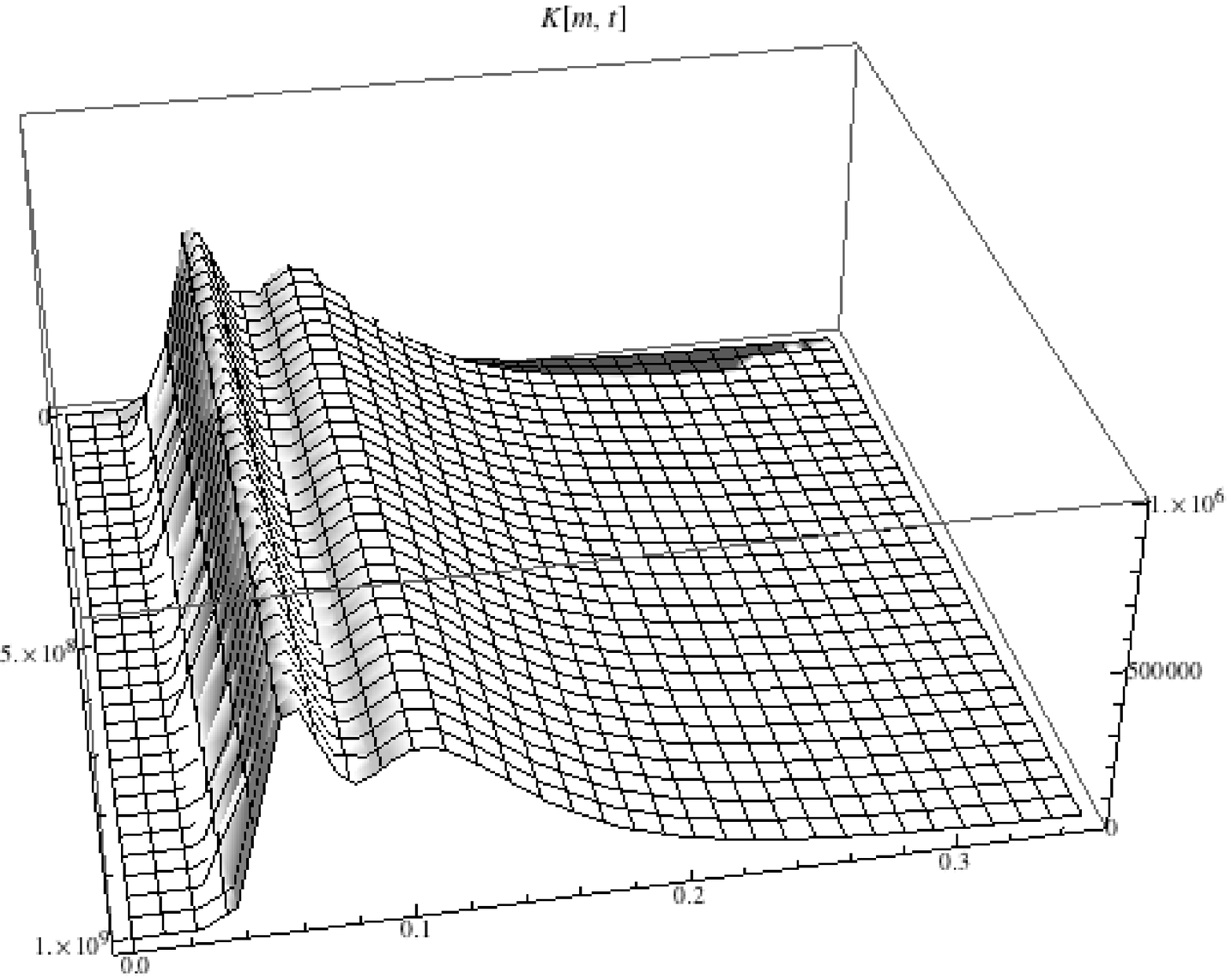}
  \end{center}
  \caption{\label{K} 
   Plot of the Kretschmann invariant $\rm R_{\mu \nu \rho \sigma} \rm R^{\mu \nu \rho \sigma}(m,r)$ 
for $m \in [0, 10^6]$, $t \in [0,2]$ $\gamma \delta \sim 1$.}
  \end{figure}

  As it is evident from Fig.\ref{K} the maximum of the Kretschmann invariant is 
  independent of the mass and it is in 
  $r_{\rm Max} \sim \sqrt{a_0}$ $(a_0 \sim l_P^2)$ localized.
  At this point we redefine the variables $t \leftrightarrow x$ 
  (with the subsequent identification $x \equiv r$)
and the metric components to bring the solution in the 
standard Schwarzschild form
\begin{eqnarray}
&& \hspace{-0.27cm} - N^2(t) \rightarrow g_{rr}(r), \nonumber \\
&& X^2(t) \rightarrow g_{tt}(r), \nonumber \\
&& Y^2(t) \rightarrow g_{\theta \theta}(r) = g_{\phi \phi}/\sin^2 \theta. 
\end{eqnarray}
Schematically the properties of the metric are the following,
\begin{eqnarray}
&& \bullet \,\,\, \lim_{r \rightarrow +\infty} g_{\mu \nu}(r) = \eta_{\mu \nu}, \nonumber \\
&& \bullet \,\,\, \lim_{r \rightarrow 0} g_{\mu \nu}(r) = \eta_{\mu \nu}, \nonumber \\
&& \bullet \,\,\, \lim_{m, a_0 \rightarrow 0} g_{\mu \nu}(r)  = \eta_{\mu \nu}, \nonumber \\
&& \bullet \,\,\, K(g) < \infty \,\, \forall r, \nonumber \\
&& \bullet \,\,\, r_{\rm Max}(K(g)) \sim \sqrt{a_0}.
\label{good}
\end{eqnarray} 
We consider the property (\ref{good}) sufficient to extend the solution in all space-time.
The solution is summarized in the following table (in the table we have not fixed the parameter $p_b^0$). 
\begin{center}
\begin{tabular}{|r|r|r|}
\hline
$ g_{\mu \nu} $& $\rm LQBH $&$ \rm Classical$\\
\hline
\hline
$g_{tt}(r)$ & 
 $\frac{\frac{(2 \gamma \delta m)^2 \Omega(\delta)}{\rho^4(\delta)} \  
                         \left(1 - \rho^2(\delta) 
\left( \frac{ 1 - \frac{2m}{r} 
\mathcal{P}(\delta)}
{1 + \frac{2m}{r} \mathcal{P}(\delta) }
\right)^2   \right) }{
\left(1 -
\left( \frac{ 1 - \frac{2m}{r} \mathcal{P}(\delta)}
{1 + \frac{2m}{r}  \mathcal{P}(\delta)}
\right)^2\right)^2
\left[\Big(\frac{\gamma \delta m p_b^{0} }{2 r}\Big)^2  + r^2\right]
}$ & $ - (1 - \frac{2m}{r})$\\
\hline
\hline
$g_{rr}(r)$ & 
 $ - \frac{\gamma^2 \delta^2 \left[ \Big(\frac{\gamma \delta m p_b^{0}}{2 r^2}\Big)^2 +1 \right]}{
  1 - \rho^2(\delta)
\left( \frac{ 1 - \frac{2m}{r} \mathcal{P}(\delta)}
{ 1 + \frac{2m}{r} \mathcal{P}(\delta)}
\right)^2   
}
$
  & $\frac{1}{1 - \frac{2m}{r}}$\\
\hline
\hline
$g_{\theta \theta}(r) $
& $\Big(\frac{\gamma \delta m p_b^{0}}{2 r}\Big)^2  + r^2$ & $r^2$\\
\hline
\end{tabular}
\label{METRICA}
\end{center}
We have said in the previous section the metric solution has two event horizons.
An event horizon is defined by a null surface $\Sigma(r,\theta) = {\rm const}.$.
The surface $\Sigma(r,\theta) = {\rm const}.$
is a null surface if the normal $n_i = \partial \Sigma/ \partial x^i$ is a null vector or
satisfied the condition $n_i n^i =0$. 
The last identity says that the vector $n^i$ is on the surface $\Sigma(r, \theta)$ itself,
in fact $d\Sigma = d x^i \partial \Sigma/\partial x^i$ and $d x^i \| n^i$.  The norm of
the vector $n_i$ is given by 
\begin{eqnarray}
n_i n^i = g^{ij} \frac{\partial \Sigma}{\partial x^i} \frac{\partial \Sigma}{\partial x^i} = 0.
\label{NullS}
\end{eqnarray}
In our case (\ref{NullS}) reduces to 
\begin{eqnarray}
g^{rr} \frac{\partial \Sigma}{\partial r} \frac{\partial \Sigma}{\partial r} +
g^{\theta \theta} \frac{\partial \Sigma}{\partial \theta} \frac{\partial \Sigma}{\partial \theta} =0.
\end{eqnarray}
and this equation is satisfied where $g^{rr}(r) = 0$ and if the surface 
is independent from $\theta$, $\Sigma(r, \theta) = \Sigma(r)$. The points where 
$g^{rr} =0$ are $r_-$ and $r_+ = 2m$.

We can write the metric in another form which is more 
similar to the Reissner-Nordstr\"om space-time.
The metric can be written in the following form 
\begin{eqnarray}
&& \hspace{-0.5cm} ds^2 = -\frac{64 \pi^2 (r - r_+) (r-r_-)(r+ r_+ {\mathcal P}(\delta) )^2 }{64 \pi^2 r^4 + a_0^2}dt^2 
\nonumber \\
&& \hspace{0.2cm} +\frac{dr^2}{\frac{64 \pi^2 (r-r_+)(r-r_-)r^4}{(r+ r_+{\mathcal P}(\delta))^2 (64 \pi^2 r^4 + a_0^2)}}
+ \Big(\frac{a_0^2}{64 \pi^2 r^2} + r^2\Big) d\Omega^{(2)},\nonumber \\
&&
\label{metricabella}
\end{eqnarray}
If we develop the metric (\ref{metricabella}) by the parameter $\delta$ 
and the minimum area $a_0$ 
at the zero order we obtain the Schwarzschild solution: 
$g_{tt}(r) = - (1 - 2m/r) + O(\delta^2)+ O(a_0^2)$, 
$g_{rr}(r) =  1/(1 - 2m/r) + O(\delta^2)+ O(a_0^2)$ and 
$g_{\theta \theta}(r) = g_{\phi \phi}(r)/\sin^2 \theta = r^2 +  O(a_0^2)$.  
We have correction to the metric from the polymer parameter $\delta$
and also from the minimum area $a_0$.

\begin{figure}
 \begin{center}
 \hspace{0.1cm}
  \includegraphics[height=6cm]{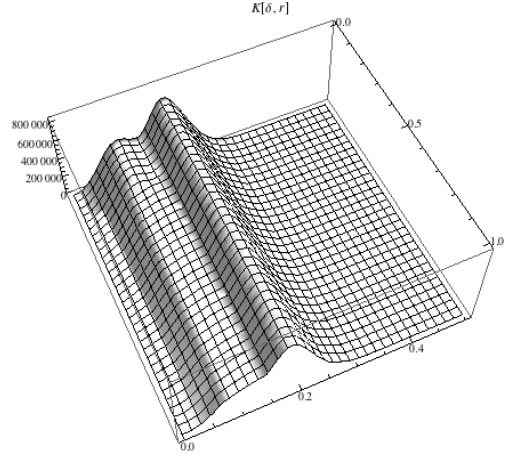}
      \end{center}
  \caption{
 Plot of the Kretschmann invariant as function of $t \in [0,0.5]$ and the 
 polymeric parameter $\delta \in[0,1]$.
  }
  \label{Kdeltat}
  \end{figure}

To check the semiclassical limit we calculate 
  the perturbative expansion of the curvature invariant for small
  $\delta$ and $a_0$ and we obtain a divergent quantity in $r=0$ at any order of
 the development.
 The regularity of $K$ is a non perturbative result, in fact for small values of the 
 radial coordinate $r$, 
$K\sim 3145728 \pi^4 r^6/a_0^4 \gamma^8 \delta^8 m^2$ diverges for $a_0\rightarrow 0$.
(For the semiclassical solution the trace of the Ricci tensor
($\rm R = R^{\mu}_{\mu}$) is not 
identically zero as for the Schwarzschild solution. We have calculated also this 
operator and we have obtained a regular quantity in $r=0$).
\begin{figure}
 \begin{center}
  \includegraphics[height=2.95cm]{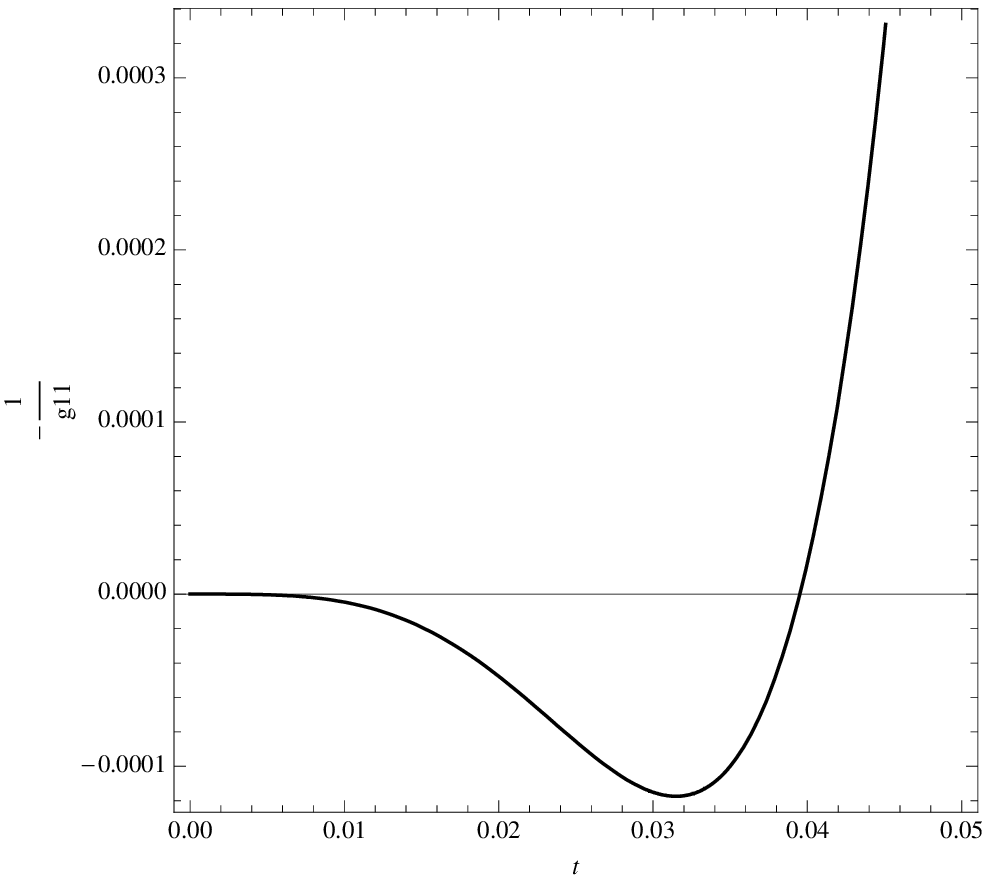}
  \includegraphics[height=2.98cm]{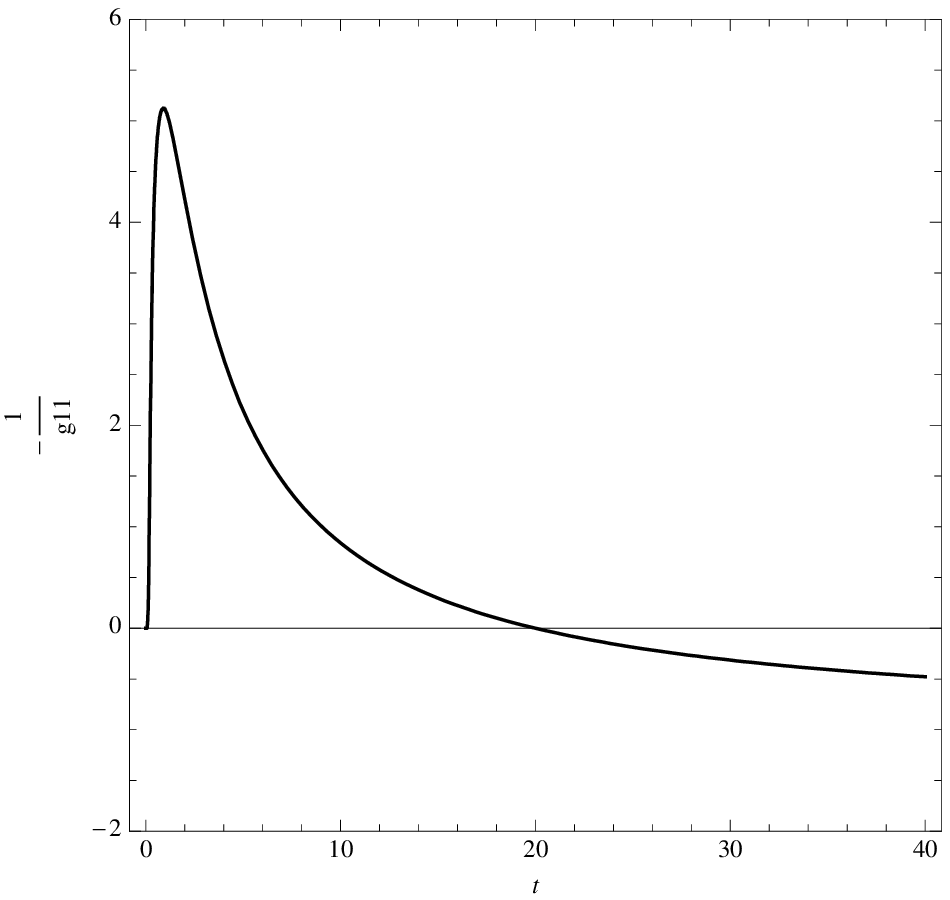}
    \end{center}
  \caption{\label{Pot2} 
 Plot of 
 $-1/g_{11}$ for $r \in [0, \sim r_-]$ (in the first picture) 
 and $-1/g_{11}$ for $r \in [\sim r_-, \infty[$ (in the second picture) .}
  \end{figure}
\begin{figure}
 \begin{center}
 \hspace{0.1cm}
  \includegraphics[height=2.95cm]{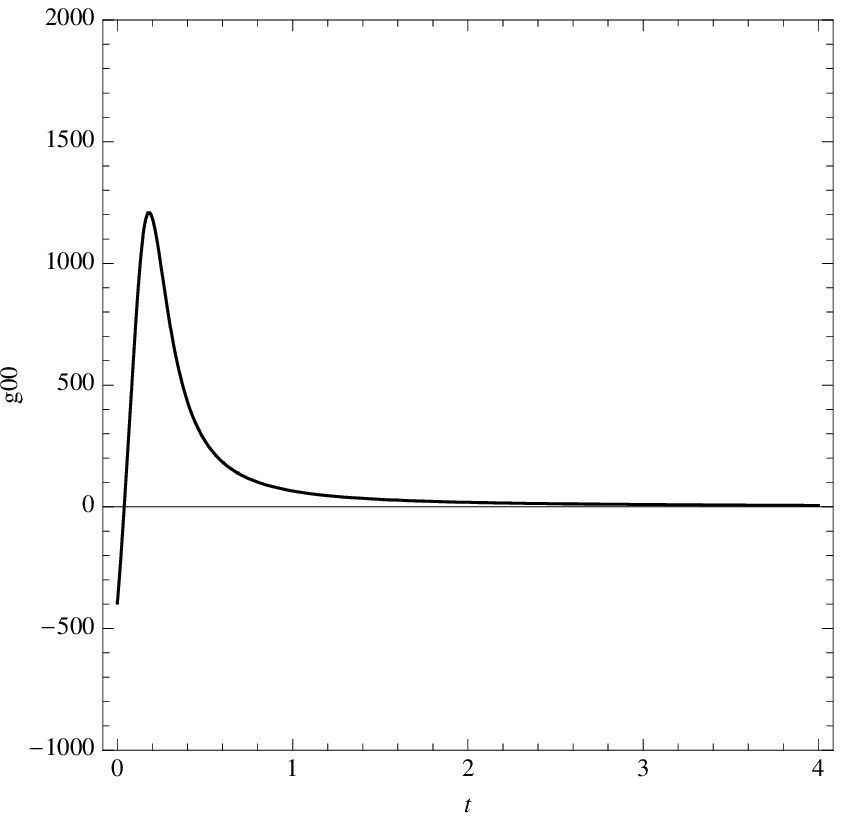}
  \hspace{0cm}
  \includegraphics[height=3cm]{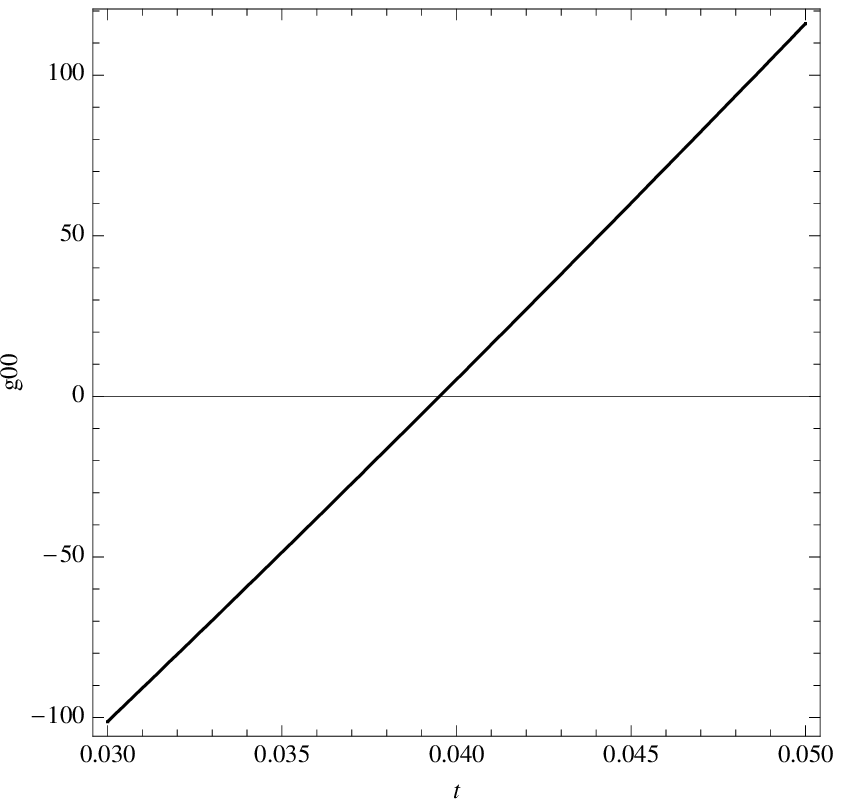}
    \end{center}
  \caption{\label{Pot1} 
 Plot of 
 $g_{00}$ for $r \in [0, \sim r_-]$ (in the first picture) 
 and $g_{00}$ for $r \in [\sim r_-, \infty[$ (in the second picture).
 For $r\rightarrow 0$ (and small $\delta$), $g_{00}\rightarrow
 -4 m^4 \pi^2 \gamma^8 \delta^8/a_0^2$.
  }
  \end{figure}
We conclude this section showing the independence of the 
pick position of Kretschmann invariant from the polymeric parameter
$\delta$. We have plotted the invariant $K(\delta,r)$ and we have 
obtained the result in Fig.(\ref{Kdeltat}).
From the picture is evident the position of the Kretschmann invariant
maximum is independent from $\delta$.
  \paragraph*{Corrections to the Newtonian potential.}
  In this paper we are interested to to singularity problem in black hole
  physics and not to the
  Post-Newtonian approximation, however we want 
  give the fist correction to the gravitational potential.
  The gravitational potential is related to the metric by 
  $\Phi(r) = -  (g_{tt}(r) + 1)/2$.
  Developing  
  the $g_{tt}$ component of the metric 
  in power of $1/r$ to the order $O(r^{-7})$, for fixed values of 
  the parameter $\delta$
  and the minimal gap area $a_0$, 
  we obtain the potential 
  \begin{eqnarray}
   && \Phi(r) = - \frac{m}{r} ({\mathcal P}-1)^2 
   - \frac{4 m^2}{r^2} {\mathcal P}({\mathcal P}^2 - {\mathcal P}+1) \nonumber \\
  && - \frac{4 m^3}{r^3}  ( {\mathcal P} -1 )^2 {\mathcal P}^2 
  +\Bigg(8 m^4 {\mathcal P}^4 - \frac{a_0^2}{128 \pi^2} \Bigg) \frac{1}{r^4}\nonumber\\
&&   +\frac{m a_0^2 ({\mathcal P} -1)^2 }{64 \pi^2 r^5} +
 \frac{m^2 a_0^2 {\mathcal P} (1 - {\mathcal P} + {\mathcal P}^2)}{16 \pi^2 r^6} + O(r^{-7}), 
 \nonumber\\
&&     \end{eqnarray}
  where ${\mathcal P} \equiv {\mathcal P}(\delta)$ is defined in (\ref{oro}).
  \section{Causal structure and Carter-Penrose diagram}\label{CPD}
  In this section we construct the Carter-Penrose diagrams \cite{Fabbri} 
  for the semiclassical metric (\ref{metricabella}). 
 To obtain the diagrams we will do many coordinate changing 
and we enumerate them from one to eight.
  
1) We can put the metric (\ref{metricabella}) 
in the form $ ds^2 = g_{00}(r(r^*)) (dt^2 - dr^{* 2})$ introducing the
  tortoise coordinate $r^*$ defined by :
  \begin{eqnarray}
  && r^* = \int \sqrt{- \frac{g_{11}}{g_{00}}} dr = \frac{1}{512 \pi^2}  
  \Big[- \frac{2 a_0^2}{ {\mathcal P}(\delta)^2 m^2 r}
  + 512 \pi^2 r \nonumber \\
  && + \frac{a_0^2 ({\mathcal P}(\delta)^2 +1)}{{\mathcal P}(\delta)^4 \, m^3} \log(r) 
  -\frac{a_0^2 +1024 \pi^2 m^4}{({\mathcal P}(\delta)^2 -1) m^3} \log|r - r_+| \nonumber \\
&&  +\frac{a_0^2  + 1024 \pi^2 {\mathcal P}(\delta)^4 m^4}{({\mathcal P}(\delta)^2 -1) {\mathcal P}(\delta)^4 \, m^3} \log |r - r_-| \Big],
  \end{eqnarray}
  
  2) The second coordinate set to use is $(u, v, \theta, \phi)$, where $u=t-r^*$ and 
  $v=t+r^*$. The metric becomes $ds^2 = g_{00}(u,v) du \, dv$.
  
  3) The singularity on the event horizon $r_+$ disappearances using the coordinates 
  ($U^+, V^+, \theta, \phi)$ defined by
   $U^+ = - \exp(- k_+ u)/k_+$,  $V^+=  \exp(- k_+ v)/k_+$, where 
    \begin{eqnarray}
   k_+ = \frac{256 \pi^2(1- {\mathcal P}(\delta)^2) m^3}{ (a_0^2 +1024 \pi^2 m^4)}.
     \label{kpiu}
  \end{eqnarray} 
   We introduce also the parametric function 
   \begin{eqnarray}
   k_-=\frac{256 \pi^2 ({\mathcal P}(\delta)^2 -1) {\mathcal P}(\delta)^4m^3}{ (a_0^2 +1024 \pi^2 {\mathcal P}(\delta)^4 m^4)}.
  \label{kmeno}
  \end{eqnarray}
  Note that $k_+>0$ and $k_-<0$.
  In those coordinates the metric is 
  \begin{eqnarray}
  && ds^2 = - \frac{64 \pi^2 (r + r_+ {\mathcal P(\delta)})^2}{64 \pi^2 r^4 + a_0^2} 
  (r - r_-)^{1- \frac{k_+}{k_-}} \nonumber \\
&&  {\rm e}^{- \frac{ k_+}{256 \pi^2} 
  \Big[- \frac{2 a_0^2}{ {\mathcal P}(\delta)^2 m^2 r}
  + 512 \pi^2 r 
   + \frac{a_0^2 ({\mathcal P}(\delta)^2 +1)}{{\mathcal P}(\delta)^4 \, m^3} \log(r) \Big]}
   dU^+ dV^+\nonumber \\
   && = - F(r)^2 dU^+ dV^+,
  \label{metricaUVpiu}
  \end{eqnarray}
  where we have introduced the function 
  $F(r)^2 = - g_{00}(r) (\partial u/ \partial U^+) (\partial v/ \partial V^+)$
  which is defined implicitly in terms of $U^+$ and $V^+$.
  
  4) Using coordinate ($t^{\prime}, x^{\prime}, \theta, \phi$) defined by
  $x^{\prime} = (U^+ - V^+)/2$, $t^{\prime} = (U^+ +V^+)/2$, the metric (\ref{metricaUVpiu})
  assumes the conformally flat form $ds^2 = F(r)^2 ( - dt^{\prime 2} + dx^{\prime 2})$. 
  In those coordinates the trajectories of constant $r$-coordinate are 
  \begin{eqnarray}
  && U^+ V^+ = t^{\prime 2} - x^{\prime 2} = -\frac{{\rm e}^{2 k_+ r^*}}{k_+^2} \nonumber \\
  && = -\frac{1}{k_+^2} (r- r_+) (r- r_-)^{\frac{k_+}{k_-}} \nonumber \\
  && {\rm e}^{\frac{ k_+}{256 \pi^2} 
  \Big[- \frac{2 a_0^2}{ {\mathcal P}(\delta)^2 m^2 r}
  + 512 \pi^2 r
  + \frac{a_0^2 ({\mathcal P}(\delta)^2 +1)}{{\mathcal P}(\delta)^4 \, m^3} \log(r) \Big]}
  \label{UpiuVpiu}
  \end{eqnarray}
 The event horizons $r_+$ and $r_-$ are localized in
  \begin{eqnarray}
  && U^+ V^+ = t^{\prime 2} -  x^{\prime 2} = 0 \, , \,\,\, r=r_+, \nonumber \\
  && U^+ V^+ = t^{\prime 2} -  x^{\prime 2} = +\infty \, , \,\,\, r=r_-.
  \label{2mrpiu}
  \end{eqnarray}  
  
\begin{figure}
 \begin{center}
  \includegraphics[height=4cm]{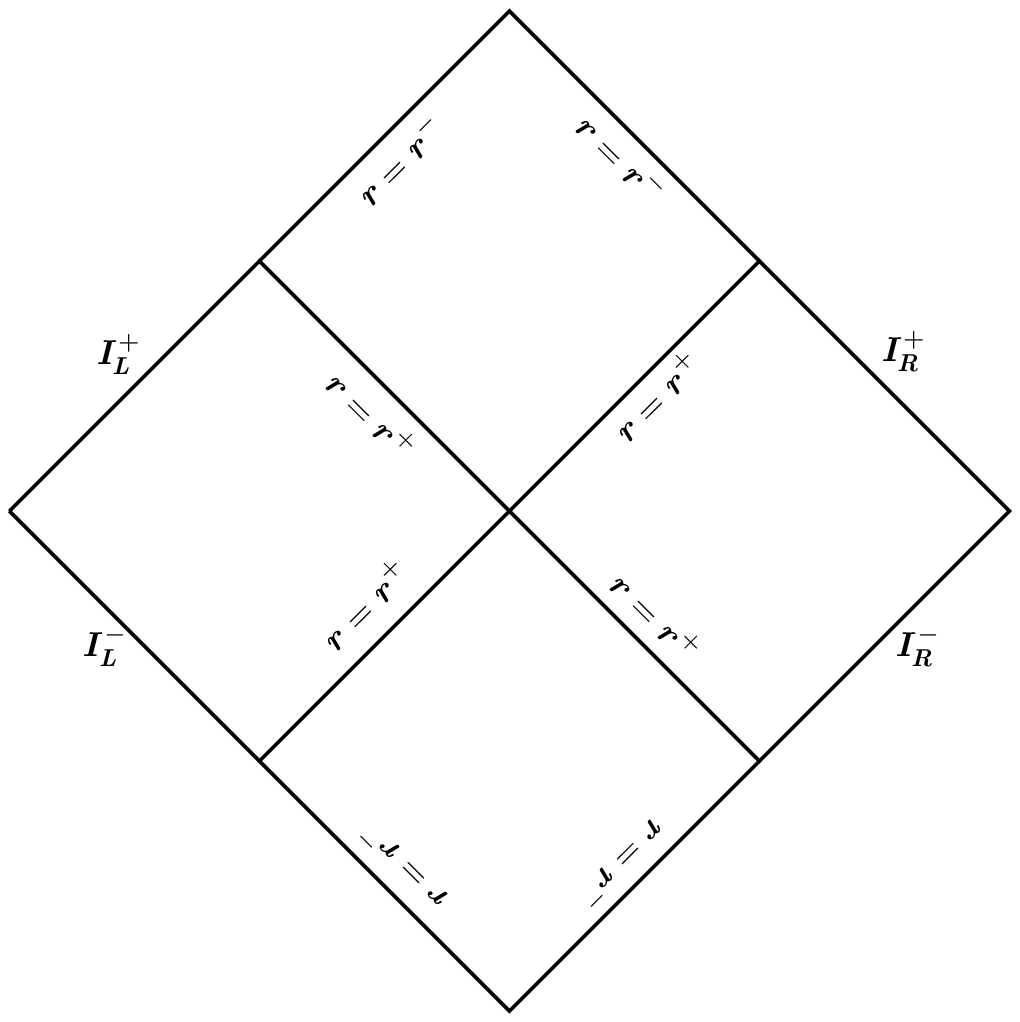}
  \includegraphics[height=4cm]{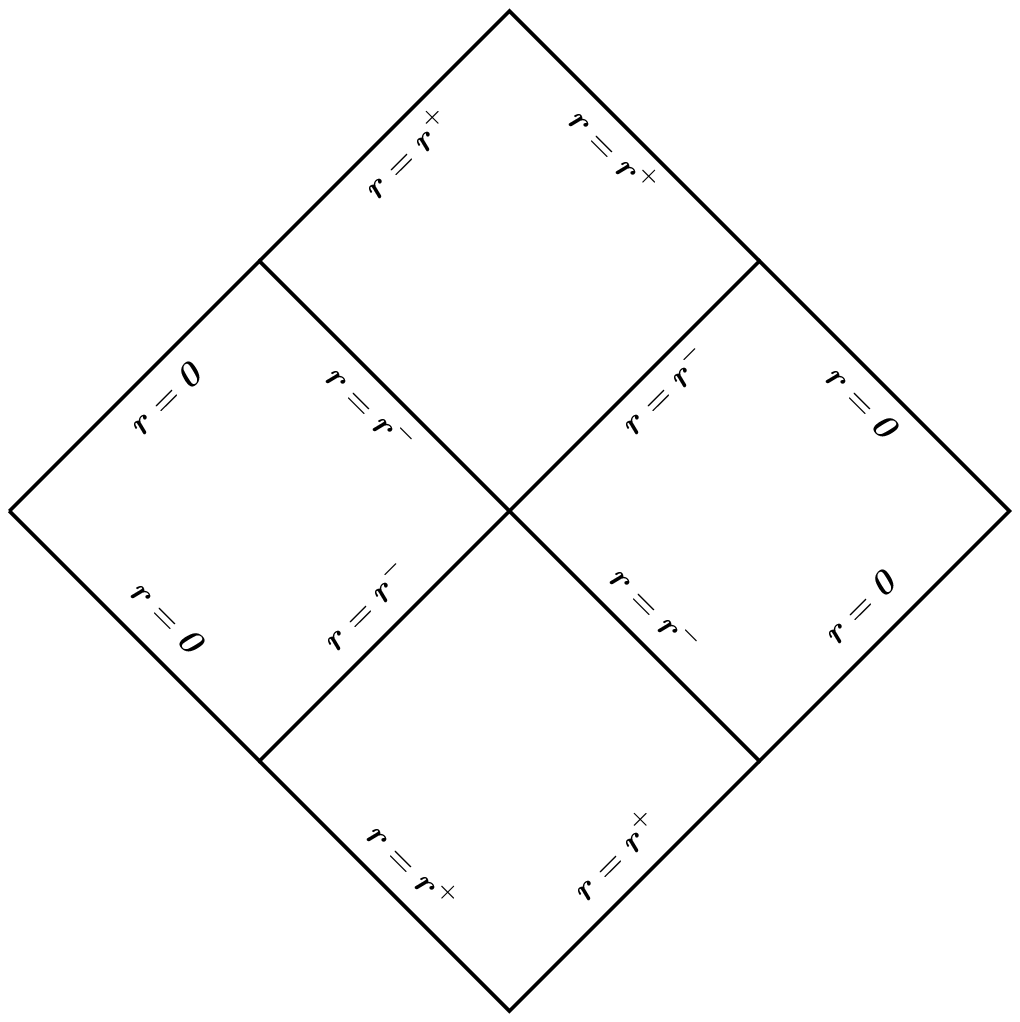}
  \end{center}
  \caption{\label{outrmeno} 
  The picture on the left represents the Carter-Penrose diagram in 
  the region outside $r_-$ and the picture on the right the diagram 
  for $r_-\leqslant r \leqslant 0$.}
  \end{figure}

  5) A first Carter-Penrose diagram for the region $r > r_-$ can be construct using coordinates 
  ($\psi, \xi, \theta, \phi$) defined by $U^+ \sim \tan[(\psi -\xi)/2]$, $V^+\sim \tan[(\psi +\xi)/2]$
  and
  $-\pi \leqslant\psi \leqslant \pi$, 
   $-\pi \leqslant \xi \leqslant \pi$ .
  The event horizon $r=r_+$ is localized in $U^+ V^+ =0$ or $\psi = \pm \xi$.
  The event horizon $r=r_-$ is localized in $U^+ V^+ = + \infty$ or:
   $\psi = \pm \xi \pm \pi$ 
  for $-\pi/2 \leqslant\xi\leqslant0$,
   $\psi = \mp \xi \pm \pi$ 
  for $0 \leqslant\xi\leqslant \pi/2$.
  The other asymptotic regions are: $I^+, I^-$ ($\psi = \mp \xi \pm \pi$),
  $i^0$ ($ \psi =0, \xi =\pi$), 
  $i^+$ ($ \psi =\pi/2, \xi =\pi/2$),
  $i^0$ ($ \psi = -\pi/2, \xi =\pi/2$).
  The Carter-Penrose diagram for this region is given in the picture on the left in 
  Fig.(\ref{outrmeno}).

6)  In the coordinates 
   introduced above, the metric (\ref{metricabella}) is not regular in $r_-$. 
  To remove the singularity in $r_-$ we introduce the coordinates 
  ($U^-, V^-, \theta, \phi$) defined by $U^- = - \exp(- k_- u)/k_-$,  $V^-=  \exp(- k_- v)/k_-$.
  In those coordinates the metric is 
  \begin{eqnarray}
  && ds^2 = - \frac{64 \pi^2 (r + r_+ {\mathcal P(\delta)})^2 }{64 \pi^2 r^4 + a_0^2} 
  (r_+ - r)^{1- \frac{k_-}{k_+}} \nonumber \\
&&  {\rm e}^{- \frac{ k_+}{256 \pi^2}  
  \Big[- \frac{2 a_0^2}{ {\mathcal P}(\delta)^2 m^2 r}
  + 512 \pi^2 r 
   + \frac{a_0^2 ({\mathcal P}(\delta)^2 +1)}{{\mathcal P}(\delta)^4 \, m^3} \log(r) \Big]}
   dU^+ dV^+\nonumber \\
   && = - F^{\prime}(r)^{2} dU^- dV^-.
  \label{metricaUVmeno}
  \end{eqnarray}
where 
  $F^{\prime}(r)^{ 2} = - g_{00}(r) (\partial u/ \partial U^-) (\partial v/ \partial V^-)$.
Now the metric is regular in $r=r_-$ but singular in $r=r^+$.
 
7) As in the region $r>r_-$ we introduce coordinates 
 ($t^{\prime \prime}, x^{\prime \prime}, \theta, \phi$) in terms of which 
 $ds^2 = F^{\prime \, 2}(r) ( -dt^{\prime \prime 2} + dx^{\prime \prime \, 2})$.
 The $r$-constant trajectories are defined by the curves 
 \begin{eqnarray}
  && U^- V^- = t^{\prime \prime \, 2} - x^{\prime \prime \, 2} = 
  \nonumber \\
  && = -\frac{1}{k_-^2} (r_- -r) (r_+ - r)^{\frac{k_+}{k_-}} \nonumber \\
  && {\rm e}^{2 \frac{ k_+}{256 \pi^2}  
  \Big[- \frac{2 a_0^2}{ {\mathcal P}(\delta)^2 m^2 r}
  + 512 \pi^2 r
  + \frac{a_0^2 ({\mathcal P}(\delta)^2 +1)}{{\mathcal P}(\delta)^4 \, m^3} \log(r) \Big]}.
  \label{UmenoVmeno}
  \end{eqnarray}
In particular the horizons $r_+, r_-$ and the point $r=0$ are defined by the curves 
\begin{eqnarray}
  && U^- V^- = t^{\prime \prime \,  2} -  x^{\prime \prime \, 2} = + \infty \, , \,\,\, r=r_+, \nonumber \\
  && U^- V^- = t^{\prime \prime \,  2} -  x^{\prime \prime \, 2} = 0 \, , \,\,\, r=r_-, \nonumber \\
    && U^- V^- = t^{\prime \prime \,  2} -  x^{\prime \prime \, 2} = - \infty \, , \,\,\, r=0.
  \label{2mrmeno}
  \end{eqnarray}  
  
  8) In coordinate ($\psi^{\prime}, \phi^{\prime}, \theta, \phi$) defined by 
  $U^- \sim \tan[(\psi^{\prime} -\xi^{\prime})/2]$, $V^+\sim \tan[(\psi^{\prime} +\xi^{\prime})/2]$.
  The event horizon $r=r_-$ is localized in $U^- V^- =0$ or $\psi^{\prime } = \pm \xi^{\prime }$,
  The event horizon $r=r_+$ is localized in $U^- V^- = + \infty$ or:
   $\psi^{\prime } = \mp \xi^{\prime } \pm \pi$ 
  for $0 \leqslant \xi^{\prime } \leqslant \pi/2$,
   $\psi^{\prime } = \pm \xi^{\prime } \pm \pi$ 
  for $0 \leqslant \xi^{\prime } \leqslant \pi/2$.
  The other asymptotic regions are defined by $r=0$ : 
  $\psi^{\prime } = \pm \xi^{\prime } \mp \pi$ for $\pi/2 \leqslant \xi \leqslant \pi$
  and 
  $\psi^{\prime } = \pm \xi^{\prime } \pm \pi$ for $- \pi \leqslant \xi^{\prime } \leqslant - \pi/2$.
  The Carter-Penrose diagram for this region is 
  the picture on the right in   
  Fig.(\ref{outrmeno}).
  
\begin{figure}
 \begin{center}
  \includegraphics[height=3cm]{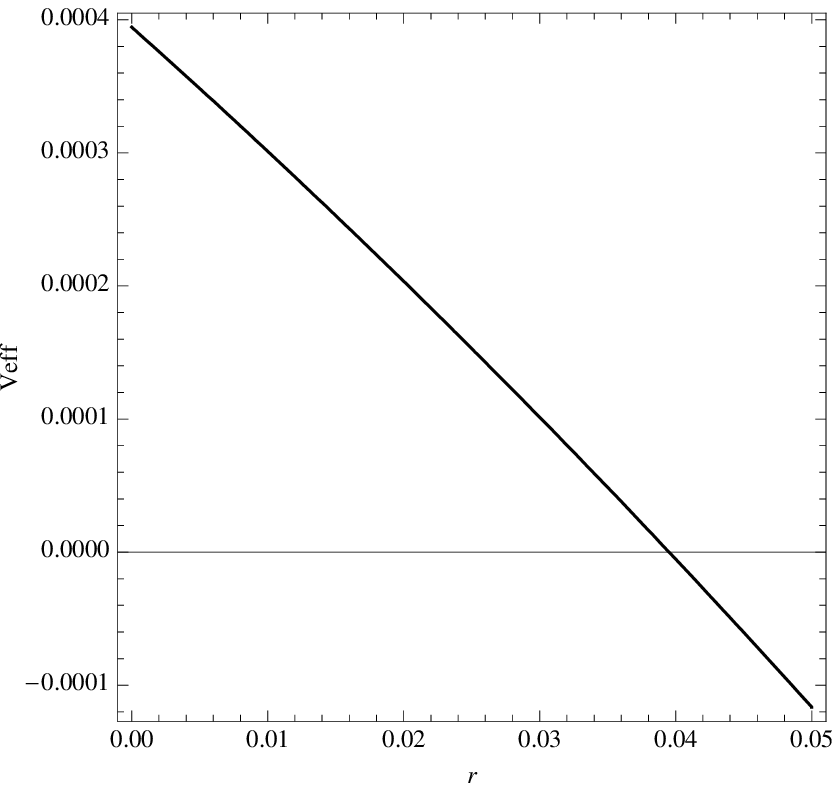}
   \includegraphics[height=3cm]{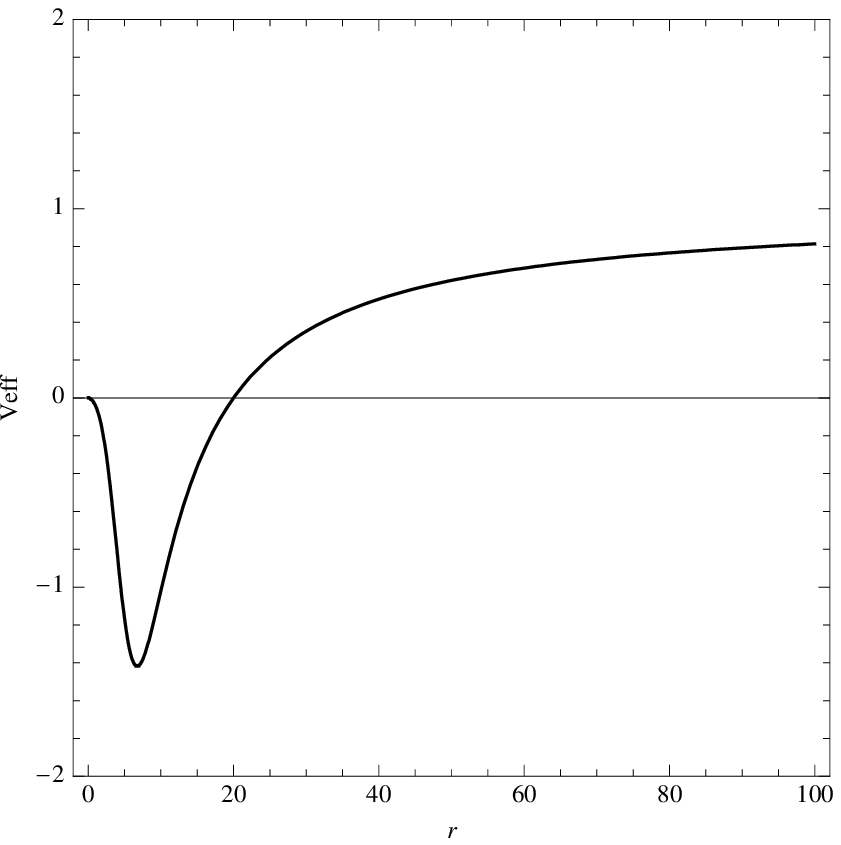}
  \end{center}
  \caption{\label{Veff} 
  Plot of $V_{eff}(r)$. On the left there is a zoom of $V_{eff}$ for 
  $r\sim0$.}
  \end{figure}

\begin{figure}
 \begin{center}
  \includegraphics[height=8cm]{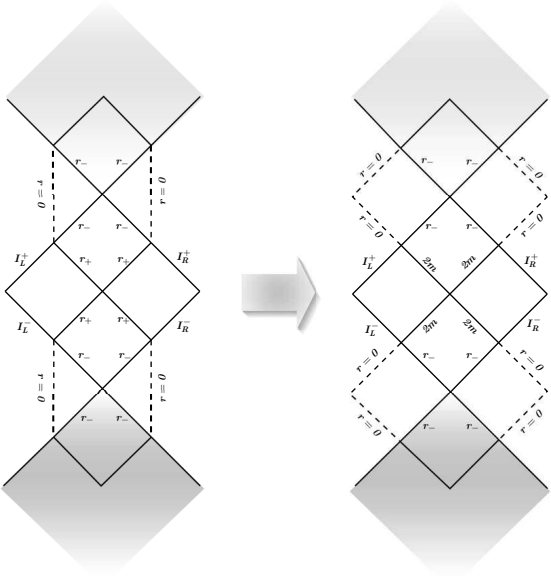}
  \hspace{1cm}
  \end{center}
  \caption{\label{penrose} 
  Maximal space-time extension of the LQBH on the right and the analog extension
  for the Reissner-Nordstr\"om black hole.}
  \end{figure}

  Now we are going to show that any massive particle could not fall in 
  $r=0$ in a finite proper time.
  We consider the radial geodesic equation for a massive point particle
  \begin{eqnarray}
 (- g_{tt} \, g_{rr}) \dot{r}^2 = E_n^2 + g_{tt},
  \label{geometricabella}  
  \end{eqnarray}
  where  ``$\,\, \dot{} \,\,$"  is the proper time derivative and $E_n$ is the point particle 
  energy. If the particle falls from the infinity with zero initial radial velocity 
  the energy is $E_n=1$. 
  We can write (\ref{geometricabella}) in a more familiar form 
    \begin{eqnarray}
   \underbrace{(- g_{tt} \, g_{rr})}_{\geqslant 0 \,\, \forall r} \dot{r}^2 + \underbrace{V_{eff}}_{-g_{tt}}(r) = \underbrace{E}_{E_n^2},
  \label{geometricabella2}  
  \end{eqnarray}
A plot of $V_{eff}$ is in Fig.(\ref{Veff}).
For $r=0$, $V_{eff}(r=0) = 4 m^4 \pi^2 \gamma^8 \delta^8/a_0^2$ 
then any particle with $E_n<V_{eff}(0)$ could not arrive in $r=0$.
If the particle energy is $E_n>V_{eff}(0)$, the geodesic equation 
for $r\sim 0$ is 
$\dot{r}^2 \sim r^4$ and integrating $\tau \sim 1/r - 1/r_0$ or 
$\Delta \tau \equiv \tau(r_0) - \tau(0) \rightarrow +  \infty$.
We can compose the diagrams in Fig.(\ref{outrmeno}) to obtain a maximal extension similar 
  to the Reissner-Nordstr\"om one, the result is represented in Fig.(\ref{penrose}).

  \section{Asymptotic Schwarzschild core near $ r \sim 0$}
  \label{core}
In this section we study the $r\sim0$ limit of the metric (\ref{metricabella}). 
If we develop the metric very closed to the point $r\sim0$ we obtain :
\begin{eqnarray}
ds^2= - (a- b \, r ) dt^2 
+ \frac{dr^2}{c \, r^4 - d \, r^5} + \frac{d \Omega}{c \, r^2}^{(2)}.
\label{metricar0}
\end{eqnarray}
The parametric functions $a, b, c, d$ are
\begin{eqnarray}
&& a= \frac{64 \Omega(\delta) m^4 \pi^2 \gamma^4 \delta^4 {\mathcal P}(\delta)^2}{a_0^2 (1+\gamma^2 \delta^2)^2}, \nonumber \\
&& b= \frac{128 \Omega(\delta) m^3 \pi^2 \gamma^2 \delta^2 {\mathcal P}(\delta)}{a_0^2 (1+\gamma^2 \delta^2)}, \nonumber \\
&& c= \frac{64 \pi ^2}{a_0^2}, \nonumber \\
&& d= \frac{128 \pi^2 (1 + \gamma^2 \delta^2) {\mathcal P}(\delta)}{a_0^2 m \gamma^2 \delta^2}.
\label{parame}
\end{eqnarray}
We consider the coordinate changing  $R= 1/r \sqrt{c}$. The point $r=0$ is mapped 
in the point $R=+ \infty$.
The metric in the new coordinates is 
\begin{eqnarray}
ds^2= - \big(1- \frac{m_1}{R} \big) dt^2 
+ \frac{dR^2}{1- \frac{m_2}{R}} + R^2 d \Omega^{(2)},
\label{metricar0b}
\end{eqnarray}
where $m_1$ and $m_2$ are functions of $m, a_0, \delta, \gamma$, 
\begin{eqnarray}
&& m_1= \frac{b}{a \sqrt{c}} = \frac{a_0 }{4 \pi m \gamma^2 \delta^2 {\mathcal P}(\delta)}, \nonumber \\
&& m_2= \frac{d}{c^{3/2}} = \frac{a_0  (1 + \gamma^2 \delta^2)}{4 \pi m \gamma^2 \delta^2 {\mathcal P}(\delta)}.
 \label{m1m2} 
\end{eqnarray}
For small $\delta$ we obtain $m_1\sim m_2$ and (\ref{metricar0b}) converges to
the Schwarzschild metric of mass 
$M \sim a_0/ 2 m \pi \gamma^4 \delta^4$. We can conclude the space-time
near the point $r\sim 0$ is described by an effective Schwarzschild metric 
of mass $M\sim a_0/m$
in the large distance limit $R\gg M$.
An observer in the asymptotic region $r=0$ experiments a Schwarzschild metric 
of mass $M\sim a_0/m$.

We now want give a possible physical interpretation of $p_b^{0}$.
If we reintroduce $p_b^{0} \sim a_0/m$ in the {\em core mass} $M$ defined above 
we obtain $M \sim p_b^{0}$,
then {\em we can interpret $p_b^{0}$ as the mass of the black hole as 
it is seen from an observer in $r\sim 0$}.
In  
 \cite{SS2} the authors 
interpret $p_b^{0}$ as the mass of a second black hole,
in our analysis instead $p_b^{0}$ 
seems to be the mass of the black hole but 
from the point of view of 
an observer in the asymptotic region $r\sim0$.


\section{LQBH termodynamics} \label{LQBHT} 
In this section we study the termodynamics of the LQBH \cite{BR}.
The form of the metric calculated in the previous section has the general form 
\begin{eqnarray}
ds^2 = - g(r) dt^2 + \frac{dr^2}{f(r)} + h^2(r) (d \theta^2 + \sin^2 \theta d \phi^2),
\label{generalmet}
\end{eqnarray}
where the functions $f(r)$, $g(r)$ and $h(r)$ depend on the mass parameter $m$
and are the components of the metric (\ref{metricabella}). 
We can introduce the null coordinate $v$ to express the metric 
(\ref{generalmet}) in the Bardeen form.
The null coordinate $v$ is defined by the relation $v = t + r^{\ast}$, where 
$r^{\ast} = \int^r d r/\sqrt{f (r) g(r)}$ and the 
differential 
is $d v = d t + d r/\sqrt{f(r) g(r)}$. In the new coordinate the metric is
\begin{eqnarray}
 d s^2 = - g(r) d v^2 + 2 \sqrt{\frac{g(r)}{f(r)}} \, d r  dv + h^2(r) d \Omega^{(2)}.
  \label{Bardeen}
\end{eqnarray}
\noindent
We can interpret our black hole solution has been generated by an effective 
matter fluid that simulates the loop quantum gravity corrections (in analogy with
the paper \cite{BR}).
The effective gravity-matter system satisfies by definition of the Einstein 
equation $G =8 \pi T$, where $T$ is the 
effective energy tensor.
The stress energy tensor for a perfect fluid compatible with the space-time
symmetries is $T^{\mu}_{\nu} = (- \rho, P_r, P_{\theta}, P_{\theta})$
and in terms of the Einstein tensor the components are 
$\rho= - G^t_t/8 \pi G_N$, $P_r = G^r_r/8 \pi G_N$
and $P_{\theta}= G^{\theta}_{\theta}/8 \pi G_N$.
The semiclassical metric to zero order in $\delta$ and $a_0$
is the classical Schwarzschild solution ($g_{\mu \nu}^{C}$) 
that satisfies $G^{\mu}_{\nu}(g^C) \equiv 0$.

\subsection{Temperature}
\begin{figure}
 \begin{center}
 \hspace{0.1cm}
  \includegraphics[height=6cm]{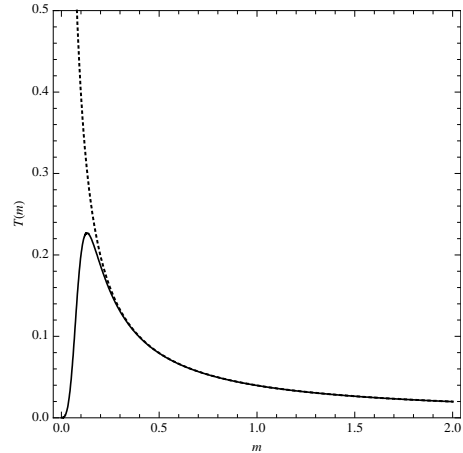}
      \end{center}
  \caption{\label{Pot} 
 Plot of the temperature $T(m)$. The continuum plot represent the LQBH temperature
 and the dashed line represent the Hawking temperature $T=1/8 \pi m$.
  }
\label{temperature}
  \end{figure}

In this paragraph we calculate the temperature
 for the quantum black hole solution and analyze the evaporation process.
The Bekenstein-Hawking temperature is given in terms of the surface gravity
$\kappa$ by $T= \kappa/2 \pi$, the surface gravity is defined by
$\kappa^2 = - g^{\mu \nu} g_{\rho \sigma} \nabla_{\mu} \chi^{\rho} \nabla_{\nu}
\chi^{\sigma}/2 = - g^{\mu \nu} g_{\rho \sigma}  \Gamma^{\rho}_{\mu 0} \Gamma^{\sigma}_{\nu 0}/2,
$
where $\chi^{\mu}=(1,0,0,0)$ is a timelike Killing vector and $\Gamma^{\mu}_{\nu \rho}$
is the connection compatibles with the metric $g_{\mu \nu}$ of (\ref{generalmet}).
Using the semiclassical metric 
we can calculate the surface gravity
in $r = 2m$ obtaining 
and then the temperature, 
\begin{eqnarray}
T(m) = \frac{128 \pi \sigma(\delta) \sqrt{\Omega(\delta)} \, m^3}{1024 \pi^2 m^4 + a_0^2}.
\label{Temperatura}
\end{eqnarray}
The temperature (\ref{Temperatura}) coincides with the Hawking temperature 
in the large mass limit. 
In Fig.\ref{temperature} we have a plot of the temperature as a function of
the black hole mass $m$.
The dashed trajectory corresponds to the Hawking 
temperature and the continuum trajectory corresponds to the semiclassical one.
There is a substantial difference for small values of the mass, in fact 
the semiclassical temperature tends to zero and does not diverge for $m\rightarrow 0$.
The temperature is maximum for $m^* = 3^{1/4} \sqrt{a_0}/\sqrt{32 \pi}$ 
and $T^*= 3^{3/4} \sigma({\delta}) \sqrt{\Omega(\delta)}/\sqrt{32 \pi a_0}$.
Also this result, as for the curvature invariant, is a quantum gravity effect,
in fact $m^*$ depends only on the Planck area $a_0$.
If we calculate the limit $\delta \rightarrow 0$ in $T(m)$ and $T^*$ we obtain
two physical quantities which are independent of $\delta$,
\begin{eqnarray}
&& \lim_{\delta \rightarrow 0} T(m) = \frac{128 \pi \, m^3}{1024 \pi^2 m^4 + a_0^2}, \nonumber \\
&&  \lim_{\delta \rightarrow 0} T^* =  \frac{3^{3/4}}{4 \sqrt{2 \pi a_0}}.
\label{TemperaturaLim}
\end{eqnarray}

\subsection{Entropy} 
\begin{figure}
 \begin{center}
  \includegraphics[height=3cm]{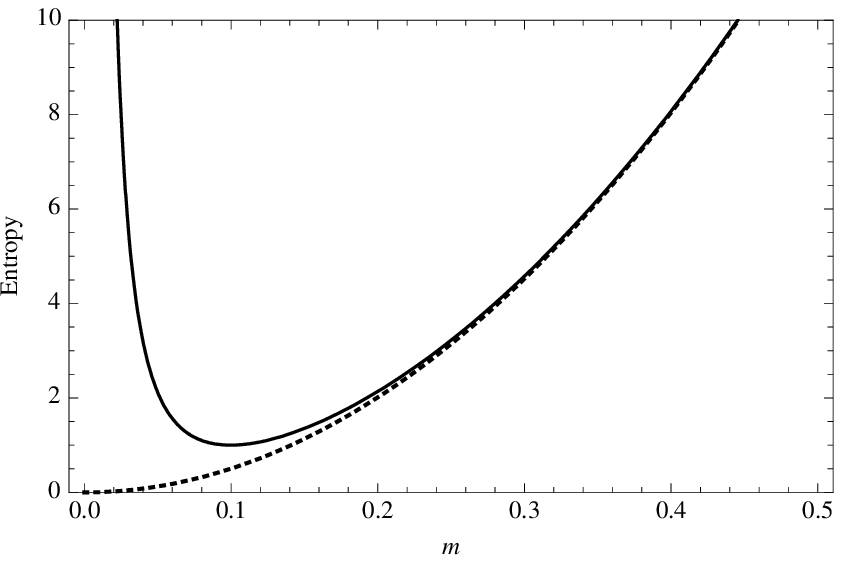}
  \end{center}
  \begin{center}
\hspace{0.001cm}  
\includegraphics[height=3.06cm]{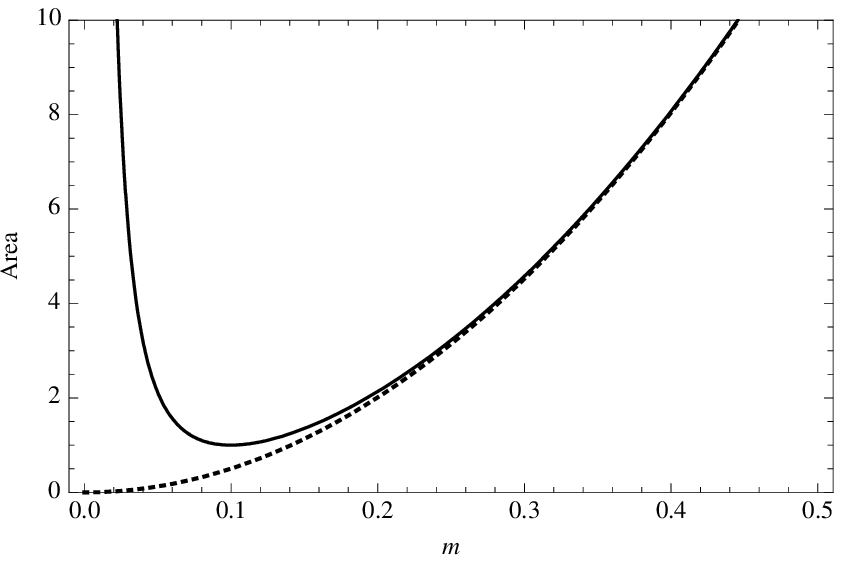}
  \end{center}
  \caption{\label{pbpc2} 
  In the first plot we have the entropy for the LQBH as function of the event horizon area 
  (dashed line represents 
  the classical area low $S_{cl} =A/4$).
  In the second plot we represent the event horizon area as function and the mass
  (dashed line represents 
  the classical area $A_{cl}=16 \pi m^2$).}
  \label{Entropy}
  \end{figure}

In this section we calculate the entropy for the LQBH metric.
By definition the entropy as function of the ADM energy is $S_{BH}=\int dm/T(m)$.
Calculating this integral for the LQBH we find 
\begin{eqnarray}
S= \frac{1024 \pi^2 m^4 - a_0^2}{256 \pi m^2 \sigma(\delta) \sqrt{\Omega(\delta)}} + {\rm const.}.
\label{entropym}
\end{eqnarray}
We can express the entropy in terms of the event horizon area.
The event horizon area (in $r=2m$) is 
\begin{eqnarray}
A = \int d \phi d \theta \sin \theta \, p_c(r)\Big|_{r = 2m} = 16 \pi m^2 +  \frac{a_0^2}{64 \pi m^2}.
\label{area}
\end{eqnarray}
Inverting (\ref{area}) 
for $m=m(A)$ and introducing the result in (\ref{entropym}) 
we obtain 
\begin{eqnarray}
 S = \frac{\sqrt{A^2 - a_0^2}}{4 \sigma(\delta) \sqrt{\Omega(\delta)}}   .
\label{entropyarea}
\end{eqnarray}
A plot of the entropy is in Fig.\ref{Entropy}.
The first plot represents entropy as a function of the event horizon area $A$.
The second plot in Fig.\ref{Entropy} represents the event horizon area as function of $m$.
The semiclassical area has a minimum value in $A=a_0$ for 
$m= \sqrt{a_0/32 \pi} $. As for the temperature
also for the entropy we can calculate the limit $\delta \rightarrow 0$
and we obtain a regular quantity which depends on the
event horizon area, on the Planck area but it is independent from $\delta$,
\begin{eqnarray}
 \lim_{\delta \rightarrow 0} S = \frac{\sqrt{A^2 - a_0^2}}{4}.
\label{entropyarealim}
\end{eqnarray}
In the limit $a_0 \rightarrow 0$, $S\rightarrow A/4$.

We want underline the parameter $\delta$ does not play any 
regularization rule in the observable quantities $T(m)$, $T^*$, $m^*$ and in the
evaporation process that we will  study in the following section.
We obtain finite quantities taking the limit $\delta \rightarrow 0$.
This is an important prediction of the model.
 \subsection{The evaporation process.}
In this section we focus our attention on the evaporation process 
of the black hole mass and in particular in the energy flux from the 
hole. First of all the luminosity can be estimated using the 
Stefan law and it is given by ${\mathcal L}(m)= \alpha A(m) T_{BH}^4(m)$,
where (for a single massless field with two degree of freedom)  
$\alpha = \pi^2/60$, $A(m)$ is the event horizon area and $T(m)$
is the temperature calculated in the previous section. 
At the first order in the luminosity the metric (\ref{Bardeen}) which incorporates 
the decreasing mass as function of the null coordinate $v$ is also a solution
but with a new effective stress energy tensor as underlined previously. 
Introducing the results (\ref{Temperatura}) and (\ref{area}) of the previous paragraphs  
in the luminosity ${\mathcal L}(m)$ 
we obtain 
\begin{eqnarray}
\mathcal{L}(m) = 
\frac{4194304 \, m^{10} \pi^3 \alpha \, \sigma^4 \Omega^2}{
(a_0^2+1024\, m^4 \pi^2)^3}.
\label{lumini}
\end{eqnarray}
Using (\ref{lumini}) we can solve the fist order differential equation
\begin{eqnarray}
- \frac{d m(v)}{d v} = \mathcal{L}[m(v)]
\label{flux}
\end{eqnarray}
to obtain the mass function $m(v)$. The result of integration with initial 
condition $m(v = 0) = m_0$ is 
\begin{eqnarray}
&& \hspace{-0.2cm}  -\frac{n_1 a_0^6+ n_2 a_0^4 m^4 \pi^2+ n_3 a_0^2 m^8 
\pi^4- n_4 m^{12} \pi^6}{n_5 m^9 \pi^3 \alpha \,
\sigma(\delta)^4 \Omega(\delta)^2} + \nonumber \\
&&  \hspace{-0.2cm} +\frac{n_1 a_0^6+ n_2 a_0^4 m^4_0+ \pi^2+ n_3 a_0^2 m^8_0 
\pi^4- n_4 m^{12}_0 \pi^6}{n_5 m^9_0 \pi^3 \alpha \,
\sigma(\delta)^4 \Omega(\delta)^2} = - v \nonumber \\
&& 
\label{v(m)}
\end{eqnarray} 
where $n_1=5$, $n_2 =27648$, $n_3 =141557760$, $n_4=16106127360$,
$n_5 = 188743680$. From the solution (\ref{v(m)})
we see the mass evaporate in an infinite time. Also in (\ref{v(m)}) we
can take the limit $\delta \rightarrow 0$ obtaining a regular quantity
independent from $\delta$. In the limit $m \rightarrow 0$ equation (\ref{v(m)}) 
becomes 
\begin{eqnarray}
   \frac{n_1 a_0^6}{n_5  \pi^3 \alpha \,
\sigma(\delta)^4 \Omega(\delta)^2 \, m^9}   =  v.
\label{v(m)2}
\end{eqnarray} 
We can take  the limit $\delta \rightarrow 0$ obtaining 
$n_1 a_0^6/n_5  \pi^3 \alpha \, m^9 \sim v$. Inverting this equation
for small $m$ we obtain: $m \sim (a_0^6/ \alpha \, v)^{1/9}$.

\section{The metric for $\delta \rightarrow 0$}
\begin{figure}
 \begin{center}
  \includegraphics[height=5cm]{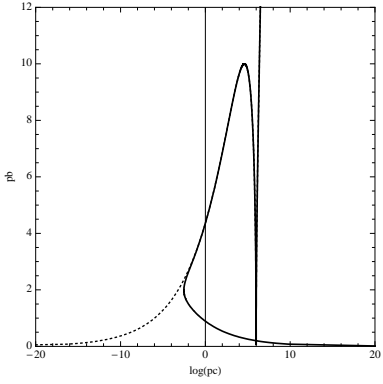}
  \end{center}
  \caption{\label{pbpcdelta0} 
  Plot ($\sqrt{|p_b^2|}/p_b^{0}, \log(p_c)$) for $\delta \rightarrow 0$.
  The dashed line represents the classical solution.}
  \end{figure}

We have shown in the previous section that same physical observable
can be defined independently from the
polymeric  parameter $\delta$. 
This result suggest to calculate 
the limit of the semiclassical metric (\ref{metricabella}) 
for $\delta \rightarrow 0$.
We will obtain a regular metric 
and we will study its space-time structure.
In the quantum theory we can not take the limit $\delta \rightarrow 0$
because we haven't weakly continuity in the polymeric parameter $\delta$.
However the LQBH'metric (\ref{metricabella}) is very close to the Reissner-Nordstr\"om metric 
which is not stable and this suggest that also (\ref{metricabella}) could be not stable
when we consider non homogeneities \cite{PInsta}. If it is the case then 
the horizon $r_-$ 
disappearances or in other words by (\ref{rmeno}), ${\mathcal P}(\delta) \rightarrow 0$.
Another motivation to calculate and to study this extreme limit of the metric is to show
that the polymeric parameter does not play any rule in the singularity problem reslution. 
For $\delta \rightarrow 0$ the ($\sqrt{|p_b^2|}/p_b^{0}/p_b^{0}, \log(p_c)$) plot is given in Fig.\ref{pbpcdelta0}. 

We redefine the metric of section (\ref{metricabella}) introducing an explicit dependence 
from $\delta$ 
(the redefinition is: $g_{\mu \nu}(r) \rightarrow g_{\mu \nu}(r; \delta)$). 
The new metric is mathematically defined by
\begin{eqnarray}
\lim_{\delta \rightarrow 0} g_{\mu \nu}(r; \delta) \equiv g_{\mu \nu}(r).
\label{limitdelta}
\end{eqnarray}

The result of this limit gives the following very simple metric which is 
independent from the polymeric parameter $\delta$,
\begin{eqnarray}
&& ds^2 = - \frac{64 \pi^2 r^3(r - 2m)}{64 \pi^2 r^4 + a_0^2} dt^2 + \frac{dr^2}{ \frac{64 \pi^2 r^3(r - 2m)}{64 \pi^2 r^4 + a_0^2}} \nonumber \\
&& \hspace{1cm} + \Big( \frac{a_0^2}{64 \pi^2 r^2} + r^2\Big) d \Omega^{(2)}.
\label{metricadelta0}
\end{eqnarray}
This metric has an event horizon in $r_+=2m$ and this is in accord 
with the solution for general values of $\delta$, in fact $\lim_{\delta \rightarrow 0} r_- = 0$.
The question now is to see if the solution is regular in all space-time and in 
particular in $r=0$.
We can calculate the Kretschmann  invariant and we obtain 
\begin{eqnarray}
&& K(r) = \frac{65536 \pi^4 r^2}{(a_0^2 + 64 \pi^2 r^4)^6}
(-6 291 456 a_0^2 \pi^6 m (2m - r) r^{12}  \nonumber \\
&& \hspace{1cm} + 50 331 648 m^2 \pi^8 r^{16} 
+ a_0^8 (15 m^2 - 24 m r + 11 r^2) \nonumber \\
&& \hspace{1cm} -128 a_0^6 \pi^2 r^4 (36 m^2 - 56 m r + 17 r^2) \nonumber \\
&& \hspace{1cm} + 4096 a_0^4 \pi^4 r^8 (294 m^2 - 272 m r + 63 r^2)).
\label{Kdeltazero}
\end{eqnarray}
The invariant (\ref{Kdeltazero}) is regular in all space-time and in particular in $r=0$.
For $a_0\sim 0$ we find $K(r) = 48 m^2/r^6 + O(a_0^2)$ and for $r\sim 0$
we have $K(r) = (983040 m^2 \pi^4 r^2)/a_0^4 + O(r^3)$ that shows the non perturbative 
character of the singularity resolution.
From the second picture in Fig.(\ref{penrose2}) is evident the $r$-coordinate of
the pick of the curvature invariant $K$ is independent from the black hole mass.

\begin{figure}
 \begin{center}
  \includegraphics[height=5cm]{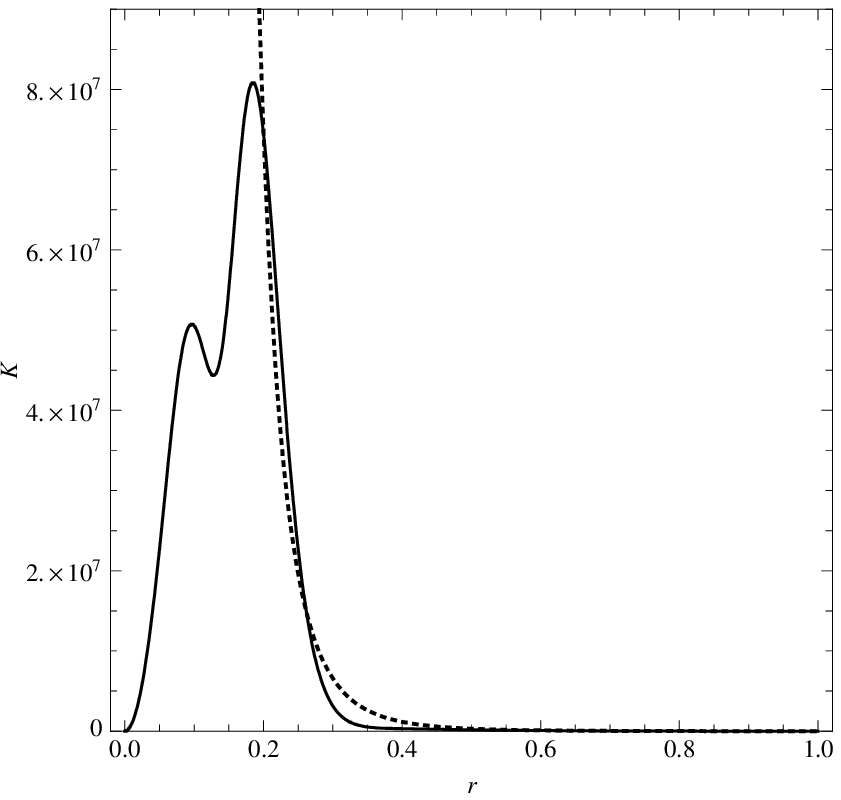}\\
 \hspace{0.5cm} \includegraphics[height=5cm]{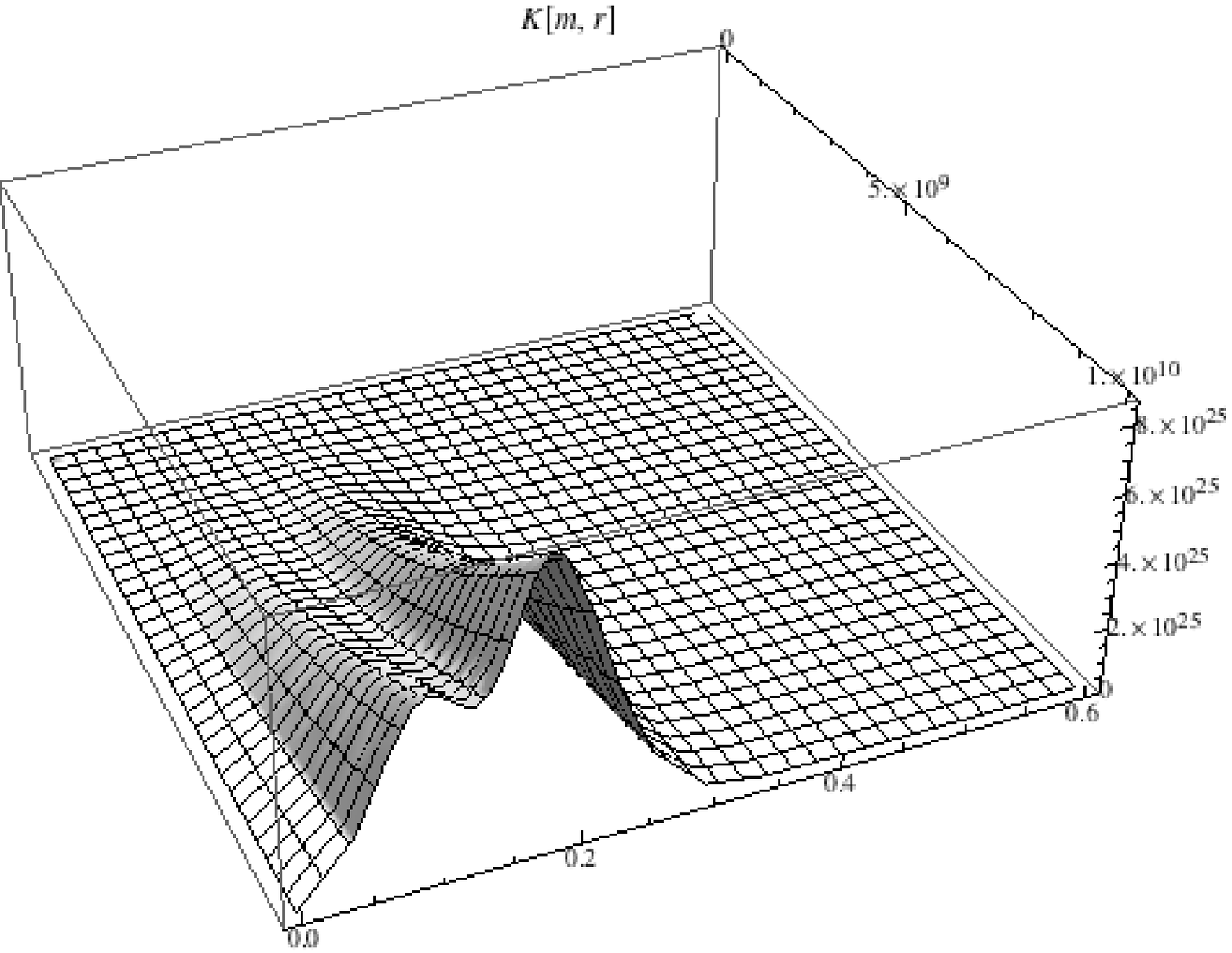}
  \hspace{1cm}
  \end{center}
  \caption{\label{penrose0} 
  Plot of the Kretschmann invariant for the metric (\ref{metricadelta0}).
  The first picture represent $K(r)$ and the second one $K(r,m)$
  for $m\in [0,10^{10}]$ and $r\in [0,0.6]$. It is manifest the position of the maximum
  of $K(m,r)$ is independent of the mass $m$.}
  \end{figure}

What about temperature, entropy and the evaporation process?
We calculate the surface gravity for the metric (\ref{metricadelta0}) and we obtain
\begin{eqnarray}
\kappa = \frac{65536 m^6 \pi^4}{(a_0^2 + 1024 m^4 \pi^2)^2}.
\label{sgdelta0}
\end{eqnarray}
This result is exactly the same quantity obtained in section (\ref{LQBHT})
but with $\delta \rightarrow 0$. From this point the analysis is the same of section
 (\ref{LQBHT}): temperature, entropy and evaporation are the same of 
 (\ref{TemperaturaLim}),
 (\ref{entropyarealim}),
 (\ref{v(m)}).

\subsection*{Causal structure and Carter-Penrose diagrams}
In this section we construct the Carter-Penrose diagrams for 
the metric obtained taking the limit $\delta \rightarrow 0$.
To obtain the diagrams we must do many coordinate changing 
and we enumerate them from one to five.

1) First of all we calculate the tortoise coordinate $r^*$ for the metric
(\ref{metricadelta0}) defined by $d r^{*2} = - g_{11}(r) dr^2/g_{00}(r)$,
\begin{eqnarray}
&& r^*= \frac{1}{64 \pi^2} \Bigg(\frac{a_0^2}{4 m r^2} +\frac{a_0^2}{4 m^2 r} + 64 \pi^2 r 
- \frac{a_0^2 \log|r|}{ 8 m^3}  \nonumber \\
&& \hspace{0.8cm} + \frac{(a_0^2 + 1024 m^4 \pi^2) \log|r-2 m|}{ 8 m^3}\Bigg).
\label{tortoise0}
\end{eqnarray}
The coordinate (\ref{tortoise0}) reduces to the Schwarzschild tortoise 
coordinate $r^*=r+2 m \log|r-2m|$ for $a_0\rightarrow 0$.
On the other side for $r\rightarrow 0$, $r^*\sim a_0/4 m r^2$.
Using coordinate $(t, r^*, \theta, \phi)$ the metric is 
\begin{eqnarray}
ds^2 = g_{00}(r(r^*))(dt^2 - dr^{*2}) + g_{\theta\theta}(r(r^*)) d \Omega^{(2)},
\label{metricstar}
\end{eqnarray}
where $g_{00}(r(r^*))$ is implicitly define by (\ref{tortoise0}) 
(from now on we will not write the $S^2$ sphere part of the metric).

2) Now we write the metric in coordinate $(v, w, \theta, \phi)$ defined
by $v= t + r^*$ and $w=t-r^*$. The metric becomes 
\begin{eqnarray}
ds^2 = g_{00}(r(r^*))dv dw = - \frac{64 \pi^2 r^3 (r - 2m)}{64 \pi^2 r^4 + a_0^2} dv dw,
\label{metricvw}
\end{eqnarray}
where $r$ is defined implicitly in terms of $v,w$.

3) We can do another coordinate changing which leaves the two space
conformally invariant. The news coordinate $(v^{\prime}, w^{\prime}, \theta, \phi)$ 
are defined by $v^{\prime}=v^{\prime}(v)$ and $w^{\prime}=w^{\prime}(w)$.
The metric is 
\begin{eqnarray}
ds^2 =  - \frac{64 \pi^2 r^3 (r - 2m)}{64 \pi^2 r^4 + a_0^2} \frac{dv}{d v^{\prime}} \frac{dw}{d w^{\prime}}
d v^{\prime} d w^{\prime},
\label{metricvwprime}
\end{eqnarray}

4) We introduce the new coordinates $(t^{\prime}, x^{\prime}, \theta, \phi)$ defined by
$t^{\prime} = (v^{\prime} + w^{\prime})/2$ and $x^{\prime} = (v^{\prime} - w^{\prime} )/2$.
The metric is 
\begin{eqnarray}
ds^2 =  \frac{64 \pi^2 r^3 (r - 2m)}{64 \pi^2 r^4 + a_0^2} \frac{dv}{d v^{\prime}} \frac{dw}{d w^{\prime}}
(-dt^{\prime 2}+dx^{\prime 2}).
\label{metricxtprime}
\end{eqnarray}
All the coordinates 
in the conformal factor 
are implicitly defined in terms of $t^{\prime}, x^{\prime}$.

At this point we choose explicitly the functions $v^{\prime}(v)$ and $w^{\prime}(w)$
to eliminate the singularity in $r=2m$.
Following the analysis of the Schwarzschild case we take $v^{\prime}(v) = \exp(v/\lambda)$
and  $w^{\prime}(w) = - \exp(-w/\lambda)$, where
$2/\lambda= 512 \pi^2 m^3/(a_0^2 +1024 \pi^2 m^4)$. 
 This is the correct coordinate changing also in
our case to eliminate the coordinate singularity on the event horizon.
We define the function $F^2(r)=-g_{00} (\partial v/\partial v^{\prime}) (\partial w/\partial w^{\prime})$
that in terms of the radial coordinate $r$ becomes 
\begin{eqnarray}
&& F^2(r)= - \lambda^2 g_{00}(r) {\rm e}^{-\frac{(v-w)}{\lambda}} =  
 - \lambda^2 g_{00}(r) {\rm e}^{-\frac{2 r^*}{\lambda}} \nonumber \\
 && = 4 \Bigg(\frac{a_0^2 + 1024 \pi^2 m^4}{512 \pi^2 m^3}\Bigg)^2 
 \Bigg(\frac{64 \pi^2 r^3}{64 \pi^2 r^4 + a_0^2}\Bigg) \times \nonumber \\
 &&\hspace{0.5cm}  \times \, {\rm e}^{- \frac{2}{\lambda} \Big[\frac{a_0^2}{256 \pi^2 m r}\Big(\frac{1}{r}+\frac{1}{m}\Big)+r
 -\frac{a_0^2}{512 \pi^2 m^3} \log(r)\Big]
 }.
\label{F}
\end{eqnarray}
The metric $ds^2= F^2(r) (-dt^{\prime 2} + d x^{\prime 2})$ is regular on the event horizon.
In the coordinates $(t^{\prime}, x^{\prime})$ the event horizon and the point $r=0$
are localized respectively in 
\begin{eqnarray}
&& t^{\prime 2} -  x^{\prime 2} =0, \nonumber \\
&& t^{\prime 2} -  x^{\prime 2} \rightarrow 2 m \exp\Big(\frac{2 a_0^2}{256 \pi^2 m \lambda r^2}\Big)
\rightarrow + \infty. \nonumber\\
&&
\label{ehs}
\end{eqnarray}

\begin{figure}
 \begin{center}
  \includegraphics[height=5cm]{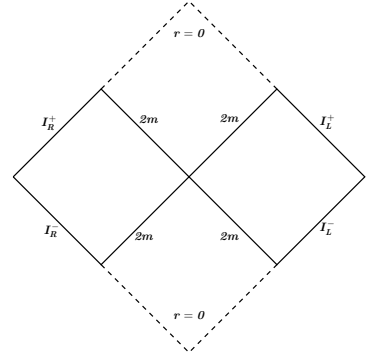}
  \hspace{1cm}
  \end{center}
  \caption{Carter-Penrose diagram for the regular space-time
  described by the metric (\ref{metricadelta0}) in coordinate 
  ($\psi, \xi$), the vertical and horizontal axes are respectively $\psi$ and 
   $\xi$. }
  \label{penrose2}
  \end{figure}

\begin{figure}
 \begin{center}
  \includegraphics[height=6cm]{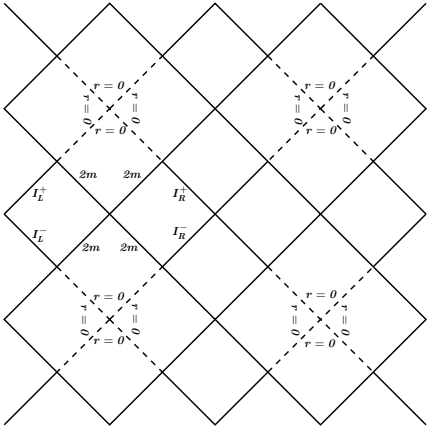}
  \hspace{1cm}
  \end{center}
  \caption{
 A possible maximal space-time extension of the Carter-Penrose diagram
  in Fig.(\ref{penrose2}).}
  \label{penrose3} 
  \end{figure}

5) We conclude writing the metric in the coordinates $(\psi, \xi, \theta, \phi)$
 defined by $v^{\prime} \sim \tan[(\psi + \xi)/2]$ and $w^{\prime} \sim \tan[(\psi - \xi)/2]$.
 The event horizon $r=2m$ is defined by the curve 
 $t^{\prime 2} - x^{\prime 2} = v^{\prime} w^{\prime} = 0$ and then by 
 the 
 $\psi = \pm \xi$.
 From (\ref{ehs}) the point $r=0$ is defined by the curve 
 $t^{\prime 2} - x^{\prime 2} = v^{\prime} w^{\prime} = +\infty$ and or by
 the segments 
 ($\psi =\mp \xi \pm \pi, \, 0 \leqslant \xi \leqslant \pi/2$), 
 ($\psi =\pm \xi \pm \pi, \, 0 \leqslant \xi \leqslant \pi/2$).
The other sectors are: $I^+, I^- $ ($\psi=-\mp\xi \pm \pi, \, -\pi \leqslant \xi \leqslant \pi$),
$i^0$ ($\psi=0, \xi=\pi$), $i^+, i^-$ ($\psi= \pm \pi/2, \xi=\pi/2$).
The Carter-Penrose diagram of the regular space-time is represented in
Fig.(\ref{penrose2}). The maximal space-time extension is represented in 
Fig.(\ref{penrose3}), the diagram can be infinitely extended in the four 
directions.

We now show that a massive particle arrives in $r=0$ 
in a finite proper time. The radial geodesic equation is 
$(dr/d \tau)^2 = E_n^2 - 1/g _{rr}$ ($\tau$ is the proper time,
$E_n$ the particle energy)
and for $r \sim 0$ reduces to
$\dot{r}(1- 64 \pi^2 m r^3/a_0^2 E_n^2)\sim - E_n$.
The $\tau(r)$ solution is $r - r_0 -16 \pi^2 m(r^4 - r_0^4)/E_n^2 a_0^2=-E_n \tau$
and the proper time to fall in $r=0$ starting from $r_0 \gtrsim 0$ is:
$\Delta \tau=\tau(0) -\tau(r_0) = (1- 16 \pi^2 mr_0^3/E_n^2 a_0^2)r_0/E_n$.
Any massive particle falls in $r=0$ in a finite proper-time interval. 

\begin{figure}
 \begin{center}
  \includegraphics[height=3.5cm]{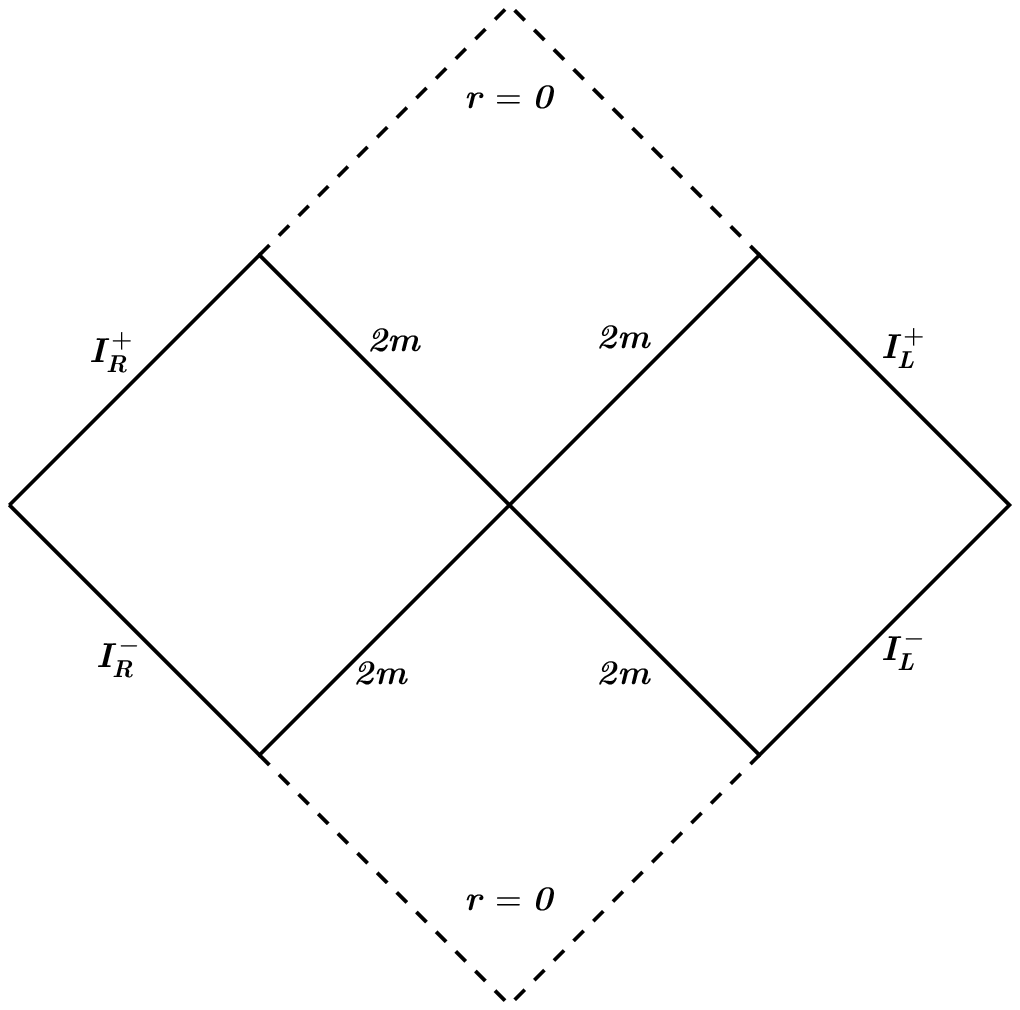}
  \includegraphics[height=3.5cm]{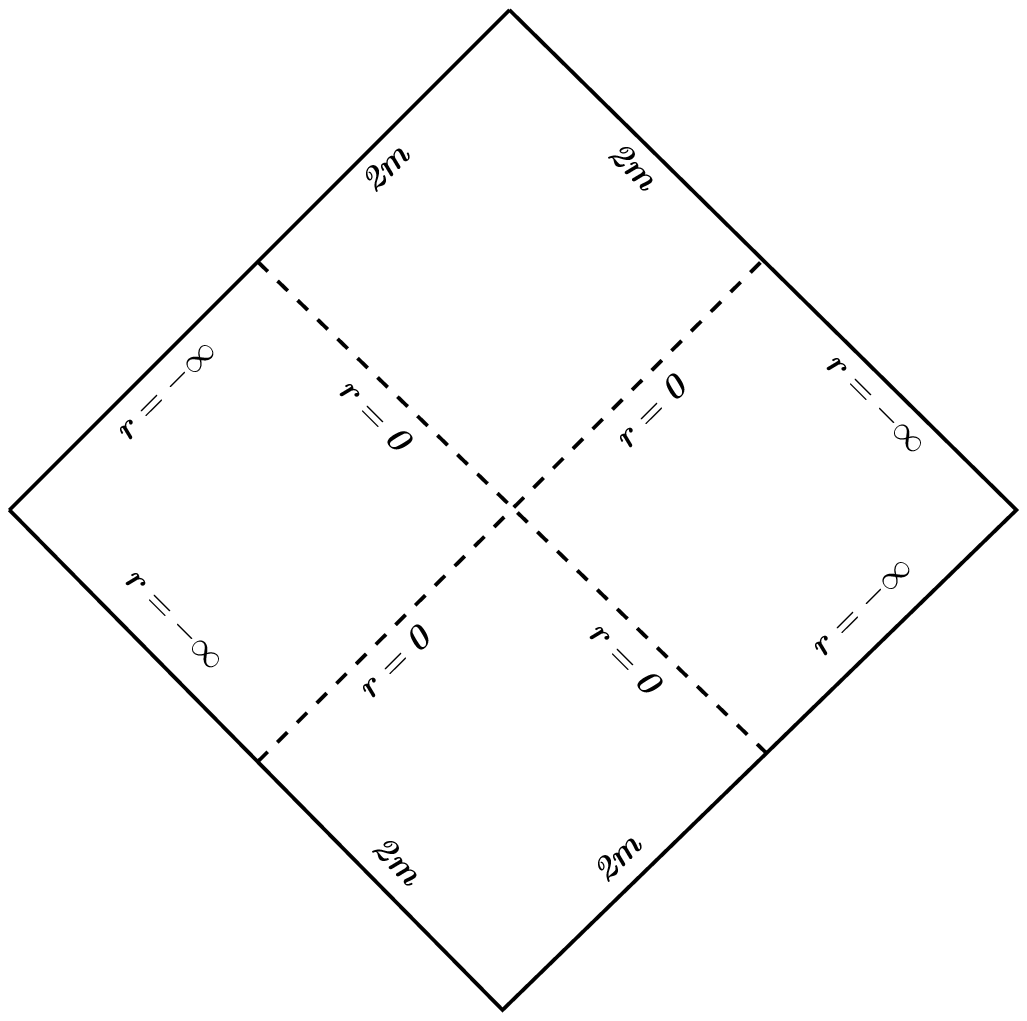}\\
  \includegraphics[height=7cm]{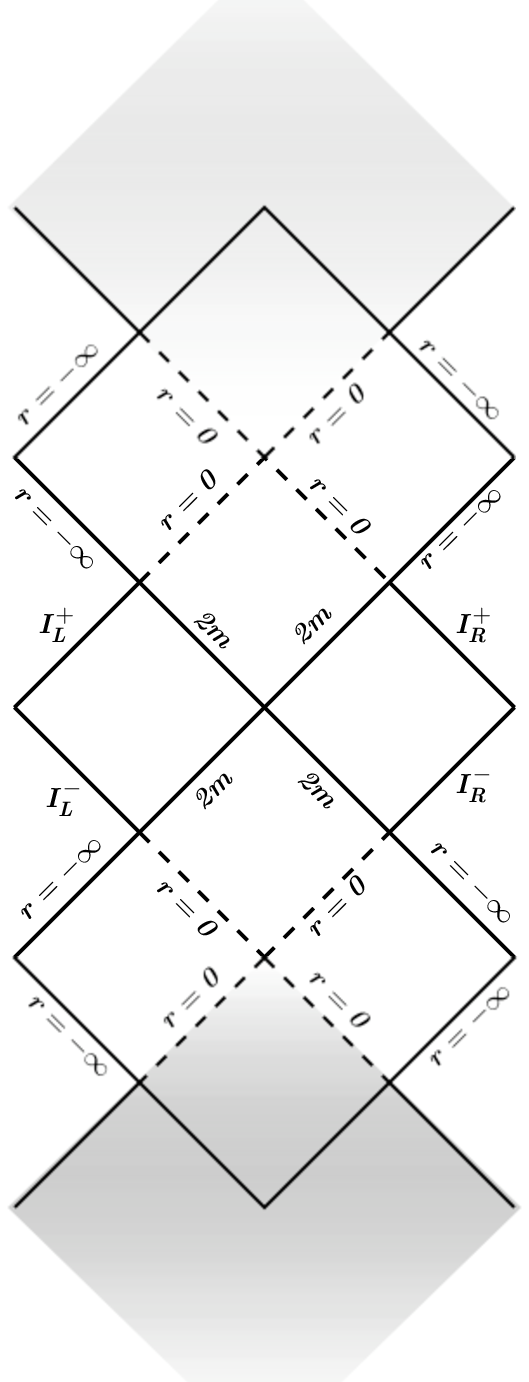}
  \end{center}
  \caption{Carter-Penrose diagrams for $r\geqslant0$ on the left and $r\leqslant0$
  on the right. The lower picture represents a maximal extension for 
  $ - \infty \leqslant r \leqslant +\infty$
  .}
  \label{penroserneg}
  \end{figure}

To conclude the analysis we extend the radial coordinate to negative values.
The surface $\Sigma(r, \theta) = r =0$ is a null surface as can be shown 
following the analysis in (\ref{LQBH}) (in particular 
$g^{rr}|_{r=0}=0$). We can extend the radial coordinate $r$ to negative values
because the space-time is singularity free. 
The metric is asymptotically flat for $r \rightarrow - \infty$ and at the order 
$O(r^{-2})$ takes the form 
\begin{eqnarray}
&& ds^2 = -\Big(1- \frac{2 m}{r} \Big) dt^2 + \frac{dr^2}{1- \frac{2 m}{r} } + r^2 d \Omega^{(2)} \, ,
\,\,\, r \leqslant0. \nonumber \\
&&
\end{eqnarray}
Because $r\leqslant0$ we have not event horizons in the negative region.
The metric (\ref{metricadelta0}) is regular in all space-time $- \infty < r <+\infty$.
\begin{figure}
 \begin{center}
  \includegraphics[height=5cm]{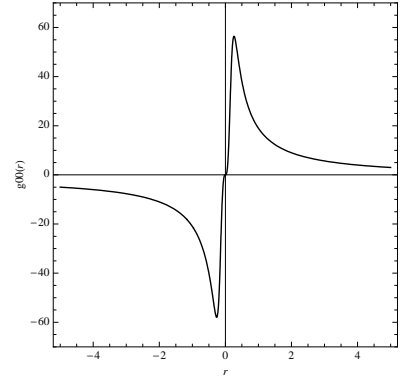}
  \hspace{1cm}
  \end{center}
  \caption{Plot of $g_{tt}(r)$ for $ - \infty < r < + \infty$. In the picture is not visible 
  the horizon in $r=2m$. }
  \label{g00all}
  \end{figure}
The Carter-Penrose diagrams are in Fig.(\ref{penroserneg}). 

We can obtain the same results of this section 
in another equivalent way.
Essentially what we have done in this section 
is to show 
that to solve the black hole singularity problem at semiclassical level
it is sufficient to replace the component $c(t)$ with the holonomy
$h=\exp(\delta c)$ without 
to replace the component $b(t)$ with the relative holonomy.
In fact the solution (\ref{metricadelta0}) can be obtained directly from the 
{\em semi-quantum} 
Hamiltonian constraint 
\begin{eqnarray}
\mathcal{C}_{sq} = - \frac{1}{2 \gamma G_N} 
\big\{ \underbrace{2 (\sin \delta c/\delta)  \  p_c}_{Quantum \,\, Sector} + 
\underbrace{(b^2 +\gamma^2)p_b/b}_{Classical \,\, Sector } \big\}.
\label{semiccla}
\end{eqnarray}
The scalar constraint (\ref{semiccla})
is classic in the $b, p_b$ sector but quantum in the $c, p_c$ sector
($N = \gamma \sqrt{|p_c|} \mbox{sgn}(p_c) / b$ and $\sigma(\delta) = 1$
). 
The constraint introduced in (\ref{CH}) is not the more general. We can introduce two different 
polymeric parameter $\delta_b$ and $\delta_c$ respectively in the directions 
$\theta, \phi$ and $r$ obtaining the constraint 
\begin{eqnarray}
&&{\mathcal C}_{\delta_b, \delta_c} = - \frac{ N}{2 G_N \gamma^2 } 
\Bigg\{ 2 \frac{\sin \delta_c c}{\delta_c} \ \frac{\sin (\sigma(\delta_b) \delta_b b)}{\delta_b} \ \sqrt{|p_c|} \nonumber \\
&& \hspace{1.0cm}+ \left(\frac{\sin^2 (\sigma(\delta_b) \delta_b b)}{\delta_b^2} + \gamma^2  \right) \frac{p_b \ \mbox{sgn}(p_c)}{\sqrt{|p_c|}} \Bigg\},
\label{cdeltabc}
\end{eqnarray}
and $N=N = \gamma \sqrt{|p_c|} \mbox{sgn}(p_c) \delta_b / \sin(\sigma(\delta_b) \delta_b b)$. 
The scalar constraint (\ref{semiccla}) is obtained taking the limit 
\begin{eqnarray}
\lim_{\delta_b \rightarrow 0} {\mathcal C}_{(\delta_b, \delta_c)}|_{\delta_c= \delta} = {\mathcal C}_{sq}.
\label{cosbc}
\end{eqnarray}

The main result is that the singularity problem is solved by a bounce of the two sphere 
on a minimal area $a_0$.
The parameter $\delta$ does not play any role in 
the singularity problem resolution. 
This is evident from the Kretschmann invariant (\ref{Kdeltazero})
which is independent from $\delta$.
The parameter $\delta$ is related to the position of the
inner horizon and for $\delta \rightarrow 0$ the horizon $r_-$ disappearances. 
\vspace{1cm}

\section*{CONCLUSIONS \& DISCUSSION}
In this paper we have
introduced a simple modification of the holonomic Hamiltonian constraint 
which gives the metric with the correct semiclassical asymptotic flat limit
when the Hamilton equations of motion are solved.
We recall here the LQBH's metric 
\begin{widetext}
\begin{centering}
\begin{eqnarray}
ds^2 = -\frac{64 \pi^2 (r - r_+) (r-r_-)(r+ r_+ {\mathcal P}(\delta) )^2 }{64 \pi^2 r^4 + a_0^2}dt^2 
+\frac{dr^2}{\frac{64 \pi^2 (r-r_+)(r-r_-)r^4}{(r+ r_+{\mathcal P}(\delta))^2 (64 \pi^2 r^4 + a_0^2)}}
+ \Big(\frac{a_0^2}{64 \pi^2 r^2} + r^2\Big) (\sin^2 \, \theta d \phi^2 + d \theta^2),
\label{metricabella2}
\end{eqnarray}
\end{centering}
\end{widetext}
We have shown the LQBH's metric (\ref{metricabella2}) has the following properties 
\begin{enumerate}
\item $ \lim_{r \rightarrow +\infty} g_{\mu \nu}(r) = \eta_{\mu \nu}$, 
\item  $ \lim_{r \rightarrow 0} g_{\mu \nu}(r) = \eta_{\mu \nu}$,
\item $\lim_{m, a_0 \rightarrow 0} g_{\mu \nu}(r)  = \eta_{\mu \nu}$,
\item $K(g) < \infty \,\, \forall r$,
\item $r_{\rm Max}(K(g)) \sim \sqrt{a_0}$.
\end{enumerate}
In particular (see point 5.) {\em the position $(r_{\rm Max})$ 
where the Kretschmann  invariant  operator 
is maximum is independent from the black hole mass and from
the polymeric parameter $\delta$}.
The metric has two event horizons that we have defined $r_+$ and $r_-$; 
$r_+$ is the Schwarzschild event horizon and $r_-$ is an inside horizon.
The solution has many similarities with the Reissner-Nordstr\"om metric
but without curvature singularities. In particular the region $r=0$ corresponds to 
another asymptotically flat region. Any massive particle can not arrive 
in this region in a finite proper time. A careful analysis shows 
the metric has a {\em Schwarzschild core} in $r\sim 0$  of mass $M\sim a_0/m$.

We have calculated the limit $
g_{\mu \nu}(\delta \rightarrow 0; r)$
of the LQBH metric
obtaining another metric regular in $r=0$. This solution can be also obtained 
from (\ref{metricabella2}) taking the limit $\delta \rightarrow 0$ or more simple
${\mathcal P}(\delta) = 0$ and $r_-=0$. The result is
\begin{widetext}
\begin{centering}
\begin{eqnarray}
 ds^2 = - \frac{64 \pi^2 r^3(r - 2m)}{64 \pi^2 r^4 + a_0^2} dt^2 + \frac{dr^2}{ \frac{64 \pi^2 r^3(r - 2m)}{64 \pi^2 r^4 + a_0^2}}
 + \Big( \frac{a_0^2}{64 \pi^2 r^2} + r^2\Big) (\sin^2 \, \theta d \phi^2 + d \theta^2).
\label{metricadelta02}
\end{eqnarray}
\end{centering}
\end{widetext}
This metric could be see as a solution of the Hamilton equation of motion for the
{\em semi-quantum} scalar constraint (\ref{semiccla}).

Our analysis shows that the 
singularity problem is solved by  
a bounce of the $S^2$ sphere on a minimum area $a_0 >0$. 
This happens for both the metrics obtained in this paper,
the first one of Reissner-Nordstr\"om type (\ref{metricabella2})
and the second one of Schwarzschild type
(\ref{metricadelta02}).
{\em The parameter $\delta$ does not play any rule in the singularity resolution problem}.
The solution (\ref{metricadelta02}) has all the good properties of (\ref{metricabella2})
and in particular it is singularity free. This metric has an event horizon in $r=2m$ and the
thermodynamics is exactly the same of (\ref{metricabella2}).
When we consider the maximal extension to $r<0$ we find a second
internal event horizon in $r=0$.

We have studied the black hole thermodynamics : temperature, entropy
and the evaporation process. 
The main results are:
\begin{enumerate}
\item The temperature $T(m)$ is regular
for $m \sim 0$ and reduces to the Bekenstein-Hawking temperature for large 
values of the mass
Bekenstein-Hawking
\begin{eqnarray}
T(m) = \frac{128 \pi  \, m^3}{1024 \pi^2 m^4 + a_0^2}.
\label{TempDis}
\end{eqnarray}
\item  The black hole entropy in terms of the event horizon area 
and the LQG minimum area eigenvalue is 
\begin{eqnarray}
 S=\frac{\sqrt{A^2 - a_0^2}}{4} 
 \label{SC}
 \end{eqnarray}
\item The evaporation process needs an infinite time in our semiclassical analysis  
but the difference with the classical result is evident only at the Planck scale.
In this extreme energy conditions it is necessary a complete quantum gravity 
analysis that can implies a complete evaporation \cite{AB}.
\end{enumerate}
We have shown it is possible to take the limit $\delta \rightarrow 0$ in 
$T(m)$, $S(A)$ and the evaporation process equation ${\mathcal F}(m; m_0, a_0) = v$
obtaining regular quantities independent of the polymeric parameter $\delta$. 
The result of the limit are physical quantities that depend only on the
Planck area and not on the polymeric parameter.

We want to conclude the discussion with a
stimulating observation.
In this paper we have calculated the temperature 
(\ref{TempDis}) that
in general we can see as a relation between  
temperature, mass and the minimum area $a_0$.
If we solve (\ref{TempDis}) for the minimum area we obtain 
the universal critical behavior $a_0 \sim (T_c - T)^{1/2}$.
The critical exponent $\zeta=1/2$ is independent from the mass and from the
particular choice of the Hamiltonian constraint modification. 
The critical temperature is the classical 
Hawking temperature $T_c=1/8 \pi m$ \cite{DO}.

\paragraph*{Some open problems.} 
In this paper we have fixed the $p_b^{0}$ parameter (which comes 
from the integration of the Hamilton equations of motion) 
introducing the minimum area $a_0$ 
(of the full theory) in the metric solution. 
In this way we have obtained a bounce of the $S^2$ sphere 
on the minimum area $a_0$. A priori it is not obvious how to obtain
the same bounce at the quantum level. 
However solving the quantum constraint 
we think we will 
obtain a bounce on a minimum area 
$a_0 \sim G_N \hbar$. 
The QEE  
contains only dimensionless quantities,
the eigenvalues $\tau, \mu$ of the operators $\hat{p}_c$, $\hat{p}_b$
and the polymeric parameter $\delta$.
When we reintroduce the length dimensions in the QEE
we have $\mu \equiv 2 p_b/\gamma l_P^2$, $\tau \equiv  p_c/\gamma l_P^2$,
then in the quantum evolution $l_P^2$
will play the rule played by $a_0$ in the semiclassical analysis 
and we will have a quantum bounce of the
wave function on $l_P^2\sim a_0$. This is manifest in the effective 
Wheeler-DeWitt equation obtained from the QEE in the limit 
$\mu \gg \delta$, $\tau \gg \delta$ \cite{work2} where $a_0^2 \sim l_P^4$ 
appears explicitly, 
\begin{eqnarray}
&& \hspace{-0.5cm} l_P^4 \Bigg(\sqrt{p_c} \frac{\partial^2 \Psi}{\partial p_b \partial p_c} + 
\frac{p_b}{4 \sqrt{pc}} \frac{\partial^2 \Psi}{\partial^2 p_b} +
\frac{1}{2 \sqrt{pc}} \frac{\partial \Psi}{\partial p_b}\Bigg) + \nonumber \\
&& \hspace{4cm} - 
 \frac{p_b}{4 \sqrt{pc}} \Psi =0. 
\label{WdW}
\end{eqnarray} 
However the quantum evolution of a coherent Schwarzschild 
state is an open problem.

A problem related to the previous one is that we have 
fixed the integration in the $x$ direction to a cell of 
finite volume ${\mathcal L}_x$ 
and this can imply a non scale invariant resolution of the singularity
problem under a rescaling ${\mathcal L}_x \rightarrow {\mathcal L}^{\prime}_x$
\cite{SScale}.

Another problem can be related to the entropy calculation.
In fact we obtain a regular entropy but we do not obtain the usual 
logarithmic correction. We think it is possible to solve this problem with 
a simple modification of the holonomic version of the Hamiltonian 
constraint or taking into account the possibility that
quantum properties of the background space-time
alter geometry near the horizon
\cite{AM}. 

\begin{figure}
 \begin{center}
  \includegraphics[height=5cm]{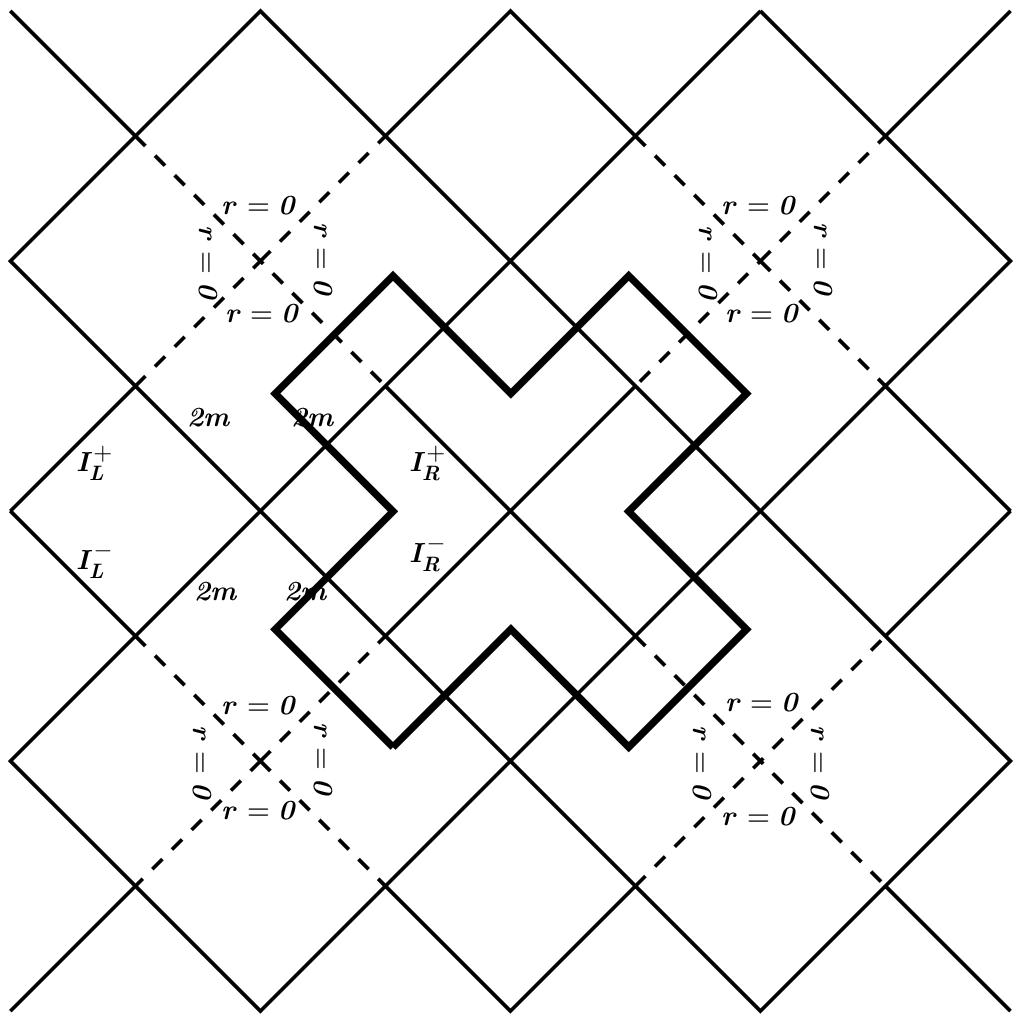}\\
  \includegraphics[height=6cm]{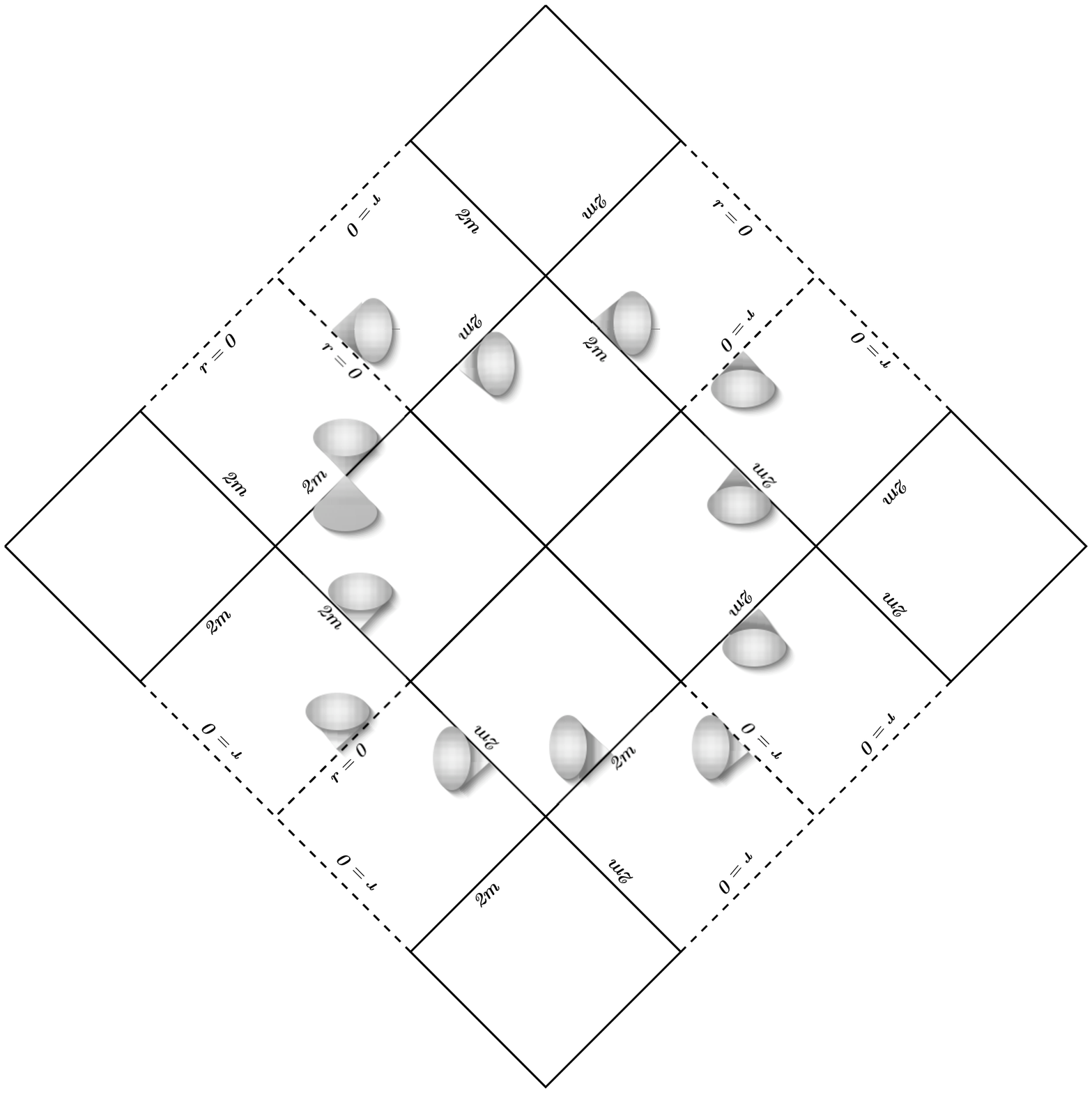}
  \end{center}
  \caption{Carter-Penrose diagram 
  of Fig.\ref{penrose3} with evidenced a light CTC curve
  in the first diagram and the light cones along a CTC curve
  in the second diagram.}
  \label{penrosectc}
  \end{figure}

Other problems could be related to the maximal extension of the space-time.
If we observe carefully the diagram in Fig.\ref{penrose3} we can see that
{\em close time-like curve} (CTC) are possible.
This is manifest in the Fig.\ref{penrosectc}
where a null CTC is represented by a close black curve.
In the second diagram of Fig.\ref{penrosectc} we have represented 
the light cones along a CTC curve.
We can have CTCs also with just one diagram if we 
identify the upper and lower extremes of the diagram 
(\ref{penroserneg}).



\section*{Acknowledgements}

We are grateful also to  Parampreet Singh, Michele Arzano and Eugenio Bianchi
for many important and clarifying discussions.

\end{document}